\documentclass{aa}  

\usepackage{graphicx}
\usepackage{txfonts}
\newcommand{\Mdot}{\ensuremath{\dot{M}}}
\newcommand{\vinf}{\ensuremath{\varv_{\infty}}}
\newcommand{\vesc}{\ensuremath{\varv_{\rm esc}}}
\newcommand{\Msol}{\text{M}\ensuremath{_{\odot}}}
\newcommand{\Rsol}{\text{R}\ensuremath{_{\odot}}}
\newcommand{\Lsol}{\text{L}\ensuremath{_{\odot}}}

\newcommand{\FW}{\textsc{\large fastwind}}
\newcommand{\MESA}{\textsc{\large mesa}}
\newcommand{\Teff}{\ensuremath{T_{\rm eff}}}
\newcommand{\logg}{\ensuremath{\log{g}}}
\newcommand{\grad}{\ensuremath{g_{\text{rad}}}}
\newcommand{\dvdr}{\ensuremath{\frac{d\varv}{dr}}}

\newcommand{\ferr}{\ensuremath{f_{\rm err}}}
\newcommand{\ferrmax}{\ensuremath{f_{\rm err}^{\text{max}}}}

\newcommand{\vturb}{\ensuremath{\varv_{\text{turb}}}}

\newcommand{\kms}{\ensuremath{{\rm km}\,{\rm s}^{-1}}}
\newcommand{\YHe}{\ensuremath{Y_{\rm He}}}
\newcommand{\Rstar}{\ensuremath{R_{\ast}}}
\newcommand{\dd}{{\rm d}}

\begin{document} 

   \title{New predictions for radiation-driven, steady-state mass-loss and wind-momentum from hot, massive stars}
   \titlerunning{Mass-loss and wind-momentum from hot, massive stars}

   \subtitle{III. Updated mass-loss rates for stellar evolution}

   \author{R. Bj\"orklund 
          \inst{1}
          \and
          J.O. Sundqvist\inst{1}\and
          S.M. Singh\inst{1} \and
          J. Puls\inst{2} \and
          F. Najarro\inst{3}}
          
   \authorrunning{R. Bj\"orklund et al.}

   \institute{KU Leuven, Instituut voor Sterrenkunde, Celestijnenlaan 200D, 3001 Leuven, Belgium\\
              \email{robin.bjorklund@telenet.be}
              \and LMU M\"unchen, Universit\"atssternwarte, Scheinerstr. 1,
     81679 M\"unchen, Germany\and Centro de Astrobiologia, Instituto Nacional de Tecnica Aerospacial, 28850 Torrejon de Ardoz, Madrid, Spain}

   \date{Received 3 August 2021; accepted 10 January 2022}

 
  \abstract
   {Massive stars lose a large fraction of their mass to radiation-driven winds throughout their entire life. These outflows impact both the life and death of these stars and their surroundings.}
   {Theoretical mass-loss rates of hot, massive stars are derived to be used in applications such as stellar evolution. The behaviour of these rates in the OB-star regime is analysed, and their effects on massive-star evolution predictions studied.}
   {Dynamically-consistent models are computed by solving the spherically symmetric, steady-state equation-of-motion for a large grid of hot, massive stars with different metallicities. The radiative acceleration is derived from non-local thermodynamic equilibrium radiative transfer in the co-moving frame, and all models cover a large spatial range from deep sub-sonic atmospheric layers into the radiation-driven and highly supersonic wind outflow. The resulting mass-loss rates are used to derive a simple scaling recipe with stellar parameters (luminosity, mass, effective temperature, metallicity), and the new recipe is used to evaluate some first impacts upon massive-star evolution tracks.}
   {We provide a new prescription for steady-state, radiation-driven mass-loss from hot, massive stars depending on their fundamental parameters. 
   In accordance with our previous work, the rates for O-stars are lower by about a factor $\sim$3 than the rates typically used in previous stellar-evolution calculations, where differences generally decrease with increasing luminosity and temperature. For cooler B-giants/supergiants we find larger discrepancies, up to one or even two orders of magnitude. This arises because we do not find any systematic increase in mass-loss rates below the so-called bi-stability region; indeed, our results do not show any sign of a significant bi-stability jump within the parameter range covered by the grid ($T_{\rm eff} \ge 15$ kK). Due to the lower mass-loss rates we find that massive-star envelopes are not easily stripped by means of standard steady-state winds, making it difficult to create classical Wolf-Rayet stars via this channel. Moreover, since the stars retain more mass right before they die as supernovae, our new rates make it possible to create black holes of higher masses than in previous models, even at Galactic metallicity. However, a remaining key uncertainty regarding these predictions concerns unsteady mass loss for very high-luminosity stars close to the Eddington limit as well as the impact of non-line-driven winds.}
   {}
   \keywords{Stars: atmospheres -- Stars: early-type -- Stars: massive -- Stars: mass-loss -- Stars: winds, outflows -- Stars: evolution}

   \maketitle
%

\section{Introduction}
    The lives of massive stars are largely controlled by their strong radiation-driven winds. The resulting mass loss has an impact on the stellar luminosity, its lifetime, apparent temperature, etc., and consequently also on the ionising radiation and UV luminosity from these massive stars \citep{Meynet94,Smith14}. These winds also shape the interstellar medium (ISM) providing chemical and mechanical feedback to their surroundings. The winds sweep up the ISM creating giant bubbles in the star's surroundings and the more massive stars even insert more energy into the ISM than supernovae \citep{Fierlinger16}. Additionally, the amount of mass loss largely determines which type of supernova (SN) the star explodes as, as well as the final mass of the remnant after that explosion \citep{Fryer01,Belczynski20}.
    
    This last effect plays a crucial role in explaining the masses of black holes as detected by gravitational waves (GW) which arise from the merging of a binary black holes. The GW detections by the Advanced-LIGO interferometers provided this first observational evidence of black hole mergers \citep{Abbott16} with black hole masses of 36 \Msol\ and 29 \Msol\ prior to merging. The surprisingly high observed masses of these black holes as well as those from more recent observations point to the need of quite weak stellar winds. Indeed, mass loss is highlighted as a main uncertainty in massive-star progenitor models of such gravitational-wave sources \citep{Abbott16b}.
    
    Accurate predictions of the mass loss of massive stars are thus necessary to predict their evolutionary paths \citep{Smith14}. Such predictions are included in stellar-evolution codes (such as the Geneva stellar evolution code \textsc{\large genec}, \citet{Eggenberger08}; \textsc{\large stern}, \citet{Heger00}, \citet{Brott11}; and \MESA, \citet{Paxton19}\footnote{These references only refer to the latest version of the code, the original work for these appear in older references.}) through simple scaling recipes depending on fundamental stellar parameters. The prescriptions used in most massive-star evolution models are typically derived from either theoretical models or empirical studies focusing on certain parts of the Hertzprung-Russel diagram (HRD), where a distinction is usually made between winds from hot stars \citep{Vink01,Nugis00,Bjorklund21} and cool stars \citep{Reimers75,deJager88,Blocker95,Kee21}. 
    
    In the first two papers of the series we developed new atmosphere and wind models in order to provide mass-loss predictions and predicted wind-structure \citep[][from now on Paper I]{Sundqvist19} and applied them to a sample of O-stars \citep[][from now on Paper II]{Bjorklund21}. The wind models solve the spherically symmetric, steady-state equation-of-motion,  relying on non-local thermodynamic (NLTE) radiative transfer in the co-moving frame to calculate the radiative acceleration throughout the atmosphere and wind outflow \citep{Puls20}. 
    In the O-star regime, the results of Paper II suggest an overall reduction in the mass-loss rate of O-stars by a factor of $\sim 2-3$ compared to the rates normally applied in stellar-evolution calculations \citep{Vink00,Vink01}, in good agreement with a range of observational results \citep[][]{Najarro11,Sundqvist11,Bouret12,Surlan13,Cohen14,Shenar15,Hawcroft21,RubioDiez21}. 
    
    In this paper we further study the dependence of the mass-loss rate on important stellar parameters, including now also the stellar mass and effective temperature on top of the stellar luminosity and metallicity. This is achieved by constructing a large grid of atmosphere and wind models that covers the O-star main-sequence and early-post main-sequence evolution of O-type stars and now also cooler B-type stars. This allows us to study the behaviour of the mass loss over the parameter ranges and to derive a new mass-loss recipe. This includes a comparison to previous predictions as well as a study of the effect of the bi-stability region (a region in effective temperature where important chemical elements that drive the wind, such as iron, change their ionisation stage) around 22 kK \citep[e.g.,][]{Vink00, Vink01, Driessen19, Krticka21}. The effect of the new mass-loss rates on massive-star evolution is analysed with a focus on the creation of Wolf-Rayet stars and high-mass black holes.
    
\section{Wind models}
    The wind-models used here rely on the interplay between the radiative transfer and the atmospheric and wind hydrodynamics. A converged model is hydrodynamically consistent which means that the one-dimensional equation of motion (eom) in the spherically symmetric, steady-state case is self-consistently solved. 
    In the case of radiation-driven stellar winds, the most important contribution to the eom comes from the radiative acceleration \grad(r) (mostly due to lines), computed for each radial point from detailed radiative transfer calculations relying on the NLTE (non-local thermodynamic equilibrium) radiative transfer of \FW\ using a co-moving frame (CMF) technique (Paper I and II, \citealt{Puls20}). 

    The initial guess of the velocity structure is taken to be a $\beta$-velocity law, where the steepness of the acceleration towards the terminal wind speed \vinf\ is set by the exponent $\beta$, which is stitched to a quasi-hydrostatic atmosphere in the inner region (see Eq. 3 of Paper II). As for \Mdot, we start from a first guess, for which we use a third of that predicted by the \citet{Vink01} recipe as justified by the results from Paper II. From test-calculations, we have verified that the exact initial conditions do not significantly affect our finally predicted model-structures; however, if these initial conditions are (very) ill-chosen this can affect the convergence behaviour of the simulation (slowing it down or even leading to non-convergence).
    During the radiative transfer the radiative acceleration is converged in an iterative process updating the occupation numbers of the included atoms. Afterwards the eom is solved by integrating in- and outwards from the sonic point from which a new velocity-structure is obtained. From the mismatch in the force balance at the sonic point, we compute a new mass-loss rate that would counter this mismatch following a basic relation $\Mdot \propto\frac{1}{\grad}$ (see Paper I). Then from $\varv(r)$ and \Mdot\ a new density structure is computed using the continuity equation. Additionally the temperature structure is updated in the same way as in Paper II (see their eqns. 6-7), 
    following \citet{Lucy71}. As also discussed in Paper II, adopting this temperature structure leads to much better and faster model-convergence, and does not introduce significant differences in the dynamics as compared to models with a temperature structure derived from flux conservation and thermal balance of electrons (see also Fig. 5 in Paper I). Throughout the iteration procedure the stellar luminosity $L_{\ast}$ and mass $M_{\ast}$ are kept fixed. The radius is updated such that the final converged model has $R_\ast \equiv r(\tau_F = 2/3)$, where $R_\ast$ is the input stellar radius and $\tau_F$ the spherically modified flux-weighted optical depth (see Paper I). All models are calculated from deep optically thick layers (lower boundary typically at a column mass of 80) to a large wind radii of about $R_{\rm max} \sim 100 R_\ast$, where the winds have almost reached their terminal wind speed $\varv_\infty \equiv \varv(r=R_{\rm max})$.

    \subsection{Convergence}
       The radiative transfer and the equation of motion influence each other, such that we need to iterate between the two steps until a converged solution is reached. As already mentioned above (see also Paper I and II), in each hydrodynamic iteration cycle the NLTE occupation numbers and the radiation field are updated until $g_{\rm rad}(r)$ is converged for the given structure. The resulting radiative accelerations and flux-weighted opacities are then used to update the velocity and temperature structures, and a new density obtained by following the procedure above for updating the mass-loss rate. This new hydrodynamic structure ($\rho(r), \varv(r), T(r)$) is then used to again converge the radiative acceleration ($g_{\rm rad}(r)$) by means of a new solution of the NLTE equations and the CMF radiative transfer. This procedure is iterated until convergence, which then also results in final convergence of the temperature structure, ionization balance and NLTE occupation numbers (the latter typically to within a maximum error of a few percent). In our hydrodynamic models final convergence is defined through the maximum error in the equation of motion (see Papers I and II). This error compares $\Gamma=\grad/g$ to the other terms in the eom according to
        \begin{align}
            \ferr(r) = 1-\frac{\Lambda(r)}{\Gamma(r)},
        \end{align}
        where
        \begin{align}
            \Lambda = \frac{1}{g}\left(\varv\dvdr\left(1-\frac{a^2}{\varv^2}\right)+g-\frac{2a^2}{r}+\frac{da^2}{dr}\right).
        \end{align}
        Our models are formally defined as converged when the maximum error radially in \ferr
        \begin{align}
            \ferrmax=\text{max}(\text{abs}(\ferr))
        \end{align}
        is less than a certain threshold throughout the complete atmosphere. Typically, we set this limit to a very stringent 2\%, however in certain regions of the HRD we allow for somewhat larger deviations (see discussion below).
        Additionally, we require both the mass-loss rate and the velocity structure not to vary more than 2\% and 3\% respectively between the last two hydrodynamic steps.
        
        \subsection{Clumping} 
        
        We have not included clumping or effects from X-rays in the models. As discussed in Paper I and Paper II, the inclusion of clumping (albeit in a highly simplified and somewhat inconsistent way, see discussion below) with an onset beyond the sonic point typically does not significantly affect the predicted mass-loss rates, because they are mostly sensitive to the conditions locally around the sonic point. This is different than, for example, models deriving a mass-loss rate only from a global energy balance constraint and not from a locally consistent solution to the equation of motion \citep{Muijres11}. Also the recent simulations by \citet{Krticka21} seem to find a general dependence on clumping, but again these models are somewhat different than ours since their Sobolev-scaled line-force shifts their wind critical point downstream from the sonic point to the supersonic parts of the outflow. By contrast, in our models the main `choke' of the wind, which effectively controls the rate of mass loss, is in principle always located in the near-sonic regions (see discussions in Papers I and II). Regarding potential inhomogeneities in sub-sonic optically thick layers, this cannot really be treated using any of the current methods for modelling `clumping' in hot-star atmospheres with winds. As such, potential effects of sub-surface structures upon line-driven mass loss are currently very difficult to assess.
        
        Moreover, let us also point out here that current inclusions of wind clumping into 1D line-driven steady-state hydrodynamic simulations (such as those presented here) are of a quite ad-hoc and somewhat inconsistent nature. For example, the standard way of introducing such wind clumping in corresponding models used for spectroscopic analysis is to assume that all wind mass is contained within optically thin clumps (see \citealt{Puls08} for a review). This can then shift the wind ionization balance and thus also have an implicit effect on $g_{\rm rad}$. But in a steady-state wind model, $g_{\rm rad}$ is rather computed for an assumed smooth (or average) wind, and not for the clumps themselves. Indeed, due to the strong inverse dependence of line-driving with density (essentially $g_{\rm rad} \sim 1/\rho^a$, where $0 < a < 1$, \citealt{Castor75}), test-calculations reveal that the clumps themselves actually do not experience significant line-driving. As such, in order to take clumping into account in hydrodynamic models one must really follow the dynamics of the (time-dependent and possibly multi-dimensional) flow in a way that accounts properly for the very different line-driving experienced by high- and low-density gas parcels. But with the elaborate techniques used to compute $g_{\rm rad}$ in this series of papers, this is not computationally possible. Instead, time-dependent line-driven wind models attempting to compute predictions for clumping \citep[e.g.,][]{Driessen19} typically use a parametrised line-force calibrated to reproduce a certain average mass loss. That is, these models are not able to predict mass-loss rates but rather use simplified radiative transfer in order to make time-dependent calculations feasible, and by means of these then predict the clumpy wind structure. Vice versa, steady-state models (such as those presented here) use very detailed techniques for computing $g_{\rm rad}$ and $\dot{M}$, on the expensive of having to assume a spherically symmetric and smooth steady-state equation of motion. Indeed, to our knowledge all non-parametrised\footnote{e.g. the attempts by \citet{Sundqvist14} to account for porosity in velocity space when computing the line-force still used a parametrised approach.} steady-state line-driven wind models  attempting to include `clumping', have computed the ionization for the clumps themselves, but then (inconsistently) solved the equation of motion for the smooth density of a steady wind (e.g., our own tests in Paper I; \citealt{Sander20}; \citealt{Krticka21}). In our opinion, it is questionable if such an approach is able to give any reasonable answers to the question if/how clumping may affect also the final (average) mass-loss rate of the star. Rather theoretical models targeting simultaneous investigations of clumping and mass-loss rates need to be carried out using time-dependent hydrodynamics, for which one then would need to develop more elaborate (but still sufficiently fast) techniques for computing $g_{\rm rad}$ than those assumed thus far in such simulations.
        
        Finally, we note that the above (of course) applies to theoretical line-driven wind simulations aiming to predict mass-loss rates. The situation is very different for diagnostic models aiming to obtain empirical mass-loss rates from direct comparison to observations, where it is crucial to properly account for the effects of wind clumping upon the diagnostic under consideration (see also discussion in Sect. 5.2.2).
        
\section{The new grid}
    \subsection{Grid setup}
\begin{figure}
            \centering
            \includegraphics[width=\hsize]{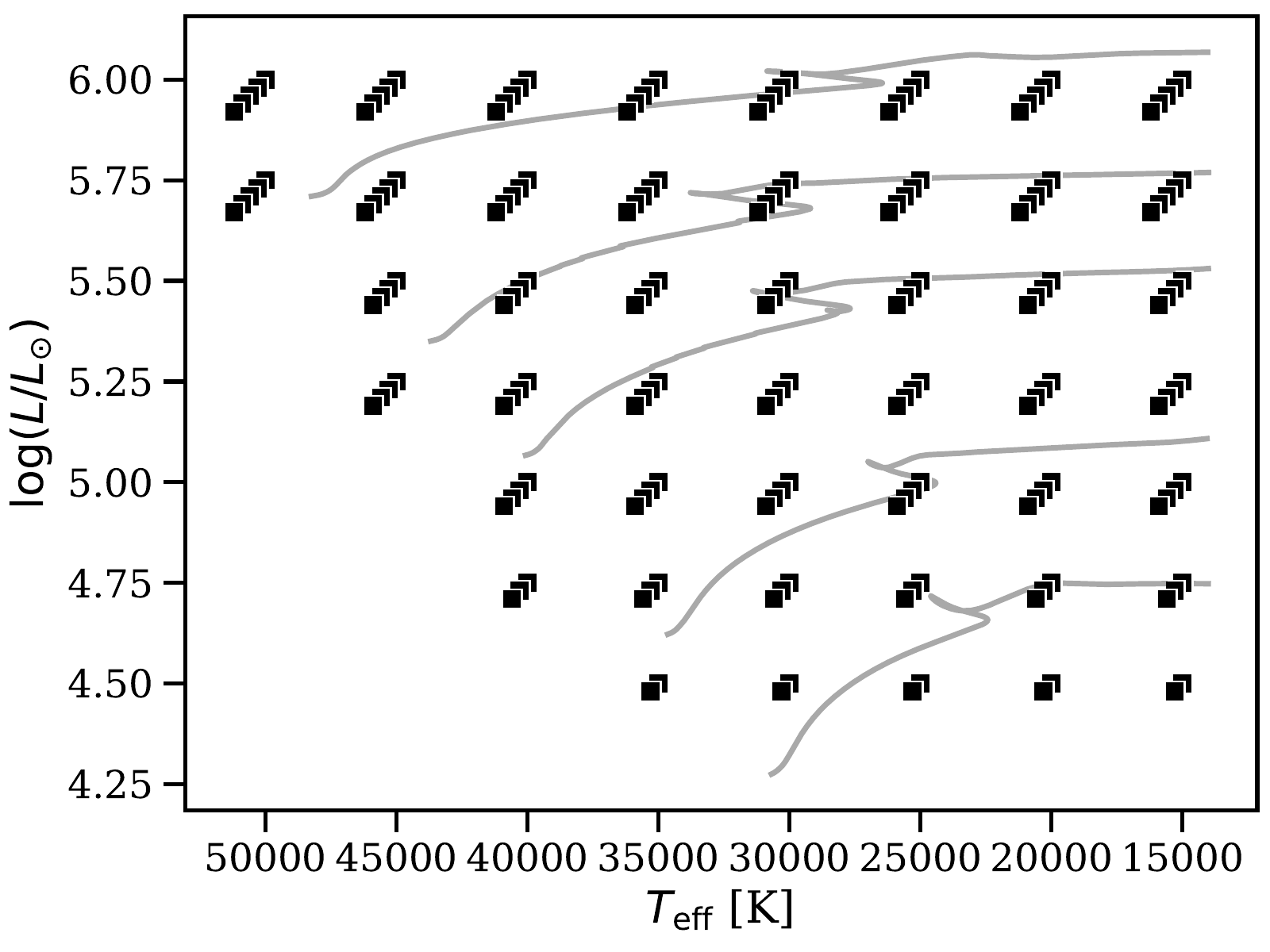}
            \caption{HR-diagram showing the coverage of the grid of models. Each point in the $L_{\ast}-\Teff$ plane represents several models with varying mass, shown by overlaying squares in the figure. Also shown are the stellar tracks including an initial rotation of 0.4 times the critical rotation for 15, 20, 30, 40 and 60 \Msol\ stars computed using \MESA.}
            \label{range}
        \end{figure}

        We set up a grid of radiation-driven wind models covering the hot region of the HR-diagram where massive stars evolve most of their lives, excluding only the cool temperatures (<15000 K), where we find, for example, red supergiants which presumably are driven by other processes \citep[e.g.][]{Kee21}, and the extremely hot temperatures of stripped Wolf-Rayet stars \citep{Sander20,Poniatowski21}. For a given metallicity $Z_\ast$, in this grid we vary the luminosity of the star, its effective temperature, and its mass. It is not a regular `cubic' grid, because we preferentially include combinations of $L_{\ast}$, \Teff, and $M_{\ast}$ that are accessible from typical predictions of stellar evolution. To set this up, we computed evolutionary tracks of stars between 15 and 80 \Msol\ using \MESA\ \citep[Modules for Experiments in Stellar Astrophysics,][]{Paxton11,Paxton13,Paxton15,Paxton18,Paxton19}. MESA is a computational tool that allows one-dimensional stellar structure and evolution models to be created for a wide variety of stellar parameters.
        The ranges covered by the grid for each parameter are $10^{4.5}-10^{6}$ \Lsol\ for $L_{\ast}$, $15000-50000$ K for \Teff, and $15-80$ \Msol\ for $M_{\ast}$. Fig. \ref{range} shows the extent of the model grid on the Hertzprung-Russel diagram. Each point represents a few models where the mass is varied according to the position on the HRD.
        Each model is calculated with a Solar metallicity of $Z_{\odot}=0.014$ and lower metallicities of $0.5 Z_{\odot}$ and $0.2 Z_{\odot}$ corresponding to those of the Large and Small Magellanic Cloud (LMC and SMC) respectively. This leads to a total of 516 wind models. This setup of the model grid allows us to investigate the dependencies on the different stellar parameters such as luminosity and metallicity, mass, and effective temperature. These latter two, as discussed in Paper II, needed a wider parameter coverage than the O-star range considered there in order to capture the proper dependence. The models all have a fixed helium number abundance $Y_{\rm He} = \frac{N_{\rm He}}{N_{\rm H}} = 0.1$ with also the micro-turbulent velocity kept fixed at \vturb\ = 10 km/s. 
        
        
    \subsection{Regions of the HR-diagram}
        The grid is chosen such that we can compute steady-state wind models that experience radiation-driven outflows. Hot, main-sequence (and post-main sequence super-giants that have not evolved too far) O-stars fall nicely within this description, and as in Paper II we find that these models normally are quite well-behaved and converge in a relatively straightforward way. There are, however, some areas of the HR-diagram where it is not so straightforward to converge the models. Two such regions are discussed below. 
     
        Several models with effective temperatures around 20 kK do not reach the strict convergence criterion  discussed above, maximum 2 \% error for the equation-of-motion over the complete radial grid, but rather show hydrodynamic solutions displaying somewhat larger maximum errors. It is important to realize, however, that even though we do not always reach this strict criterion, this does not mean that the models in general also fluctuate significantly in their predicted mass-loss rates; rather this reflects the general difficulty of converging the full equation-of-motion in these types of steady-state line-driven wind models (see also discussions in \citealt{Sander17}, \citealt{Krticka10}, \citealt{Gormaz21}). Indeed, in many of these iteration steps the mass-loss rate changes only marginally and \ferrmax\ < 10\%, with most of the atmosphere showing significantly lower errors. As such, we are still able to use these results for the wind structure and mass-loss rate (actually, our 2\% base-criterion is very stringent, e.g. the alternative CMF O-star model by Sander et al. (2017) applied a 5\% criterion and also allowed for larger deviations close to the inner and outer boundaries). In models which do not reach this strict criterion then, we calculate the final mass-loss rate as an average of all \Mdot\ in the hydrodynamic iteration steps with \ferrmax<10\%, i.e. we compute $\langle \dot{M}_{<10 \%} \rangle$. To justify this averaging-approach, we compare the predicted mass-loss rate of all formally converged models, i.e. those which reach $\dot{M}_{<2 \%}$, to the \Mdot\ that would have been derived if we instead had applied the averaging-method $\langle \dot{M}_{<10 \%} \rangle$ for also these model-stars. From this, we find that the difference $|\frac{\Mdot_{<2 \%}-\langle \Mdot_{\rm <10\%} \rangle}{\Mdot_{<2 \%}}|$ is generally very small, <5\%; indeed, in our complete grid we only find three outliers, for which the difference is then still a modest $\sim$20\%. This comparison supports the use of $\langle \Mdot_{\rm <10\%} \rangle$ as the mass-loss rate of models that do not formally reach $\Mdot_{<2 \%}$. 
    
        We limit the parameter-range of the grid to a maximum luminosity $\log L_{\ast}/\Lsol = 6$. Above this limit, the strong radiation field often induces a radiation force that exceeds gravity already in the dense, sub-surface regions of the star. Such deep-seated outflows cannot be well modelled with our current technique, both because our atomic data base does not contain the high ionisation stages that are needed in these hot atmospheric layers, and because a steady-state approach there becomes highly questionable \citep[e.g.][]{Jiang18,Schultz20}. For now we therefore do not include very high-luminosity regions in our model grid, presumably involving stars that would spectroscopically be classified as hydrogen-rich Wolf Rayet stars (WNh, sometimes called `slash stars' or `O-stars on steroids') and luminous blue variables (LBV). 

\section{The new recipe}\label{sec:recipe}
\begin{figure}
        \centering
        \includegraphics[width=\hsize]{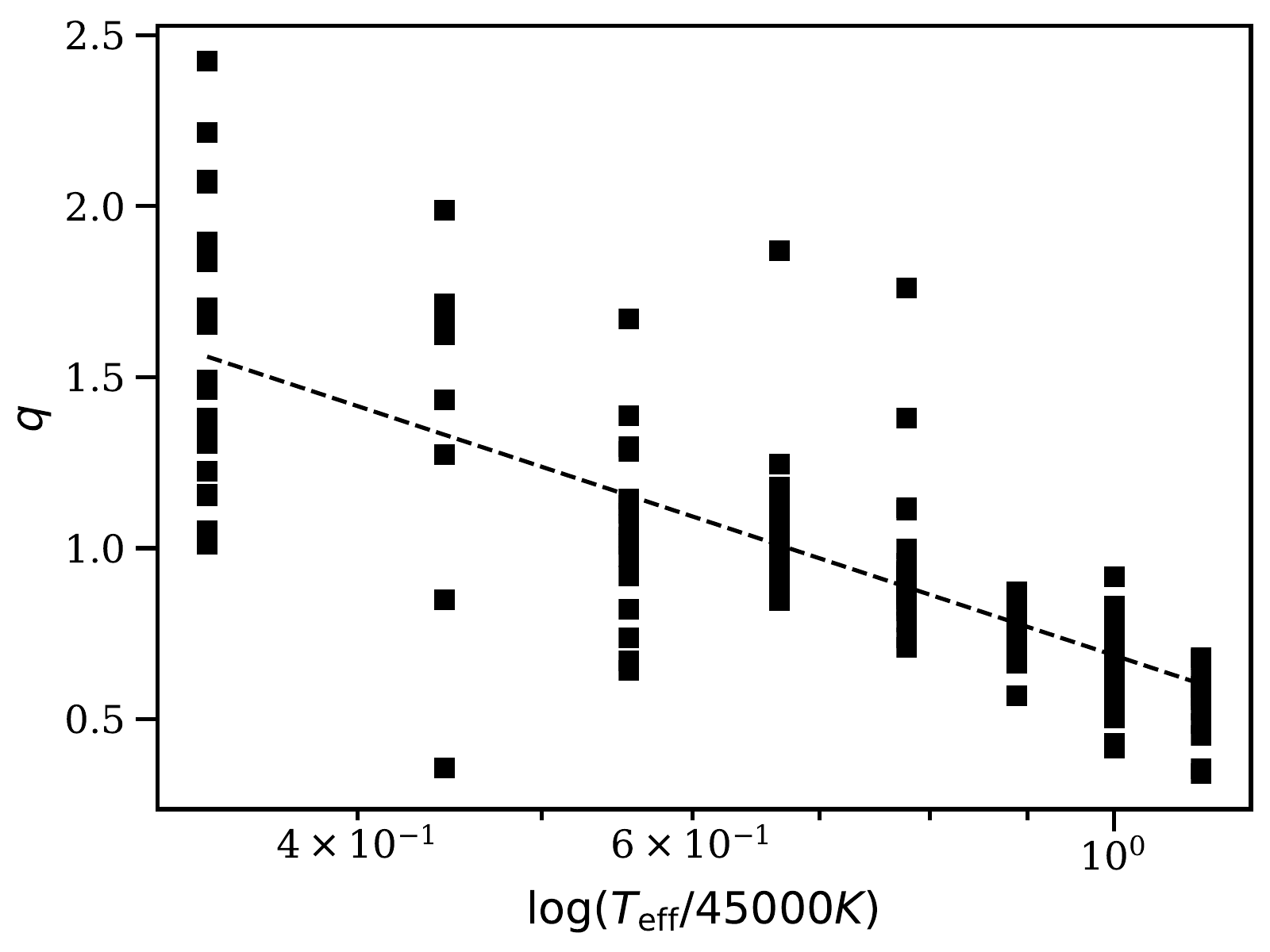}
        \caption{Dependence of the exponent $q$ on $\log\Teff$ including a linear fit with $\log\Teff$.}
        \label{q}
    \end{figure}

    The newly computed atmosphere and wind models allow us to analyse trends of the mass-loss rate with the stellar parameters that are varied in the grid. To this end, we aim to provide a simple theoretical prescription of \Mdot\ that depends on the stellar luminosity, mass, effective temperature and metallicity. For this we assume a form: 
    \begin{align}
        \Mdot \propto L_{\ast}^m M_{\rm eff}^n \Teff^p Z_{\ast}^q,
    \end{align}
    that is 
    \begin{align}\label{mnpqC}
        \log\Mdot = C + m\log L_{\ast} + n\log M_{\rm eff}  + p\log \Teff  + q\log Z_{\ast},
    \end{align}
    where $M_{\rm eff} = M_{\ast}(1-\Gamma_e)$ is the effective stellar mass, reduced by the electron-scattering Eddington parameter $\Gamma_e = \frac{\kappa_e L_{\ast}}{4\pi c G M_{\ast}}$ applying a constant $\kappa_e$ in the model-fit. The value for $\kappa_e$ is calculated by
    \begin{align}
        \kappa_e = \frac{0.4\left(1+i_{\rm He} Y_{\rm He}\right)}{1+4Y_{\rm He}},
    \end{align}
    with $i_{\rm He}$ the ionisation stage of helium (with $i_{\rm He} = 0$ for neutral Helium). For O-stars, with doubly-ionised helium and our assumed $N_{\rm He}=0.1$, this gives the typical value of $0.34 \rm \ cm^2/g$.
    Using the mass-loss rates obtained in our grid we can fit Eq. \eqref{mnpqC} and derive the best matching, $m$, $n$, $p$, $q$ and $C$ coefficients. As was also found in Paper II, though, one value for the exponent $q$ (metallicity-dependence) does not properly represent the grid results, because the parameter itself depends quite strongly on the other stellar parameters. In Paper II we showed that a linear dependence with $\log L_{\ast}$ can capture the variation of the metallicity dependence across the grid. Having access now to a wider range of parameters for the wind models, this variation is characterised more properly by using $\log\Teff$ instead of $\log L_{\ast}$. In the previous grid of O-stars, this was however essentially the same because of the strong relationship of $L_{\ast}$ and \Teff\ for models within that O-star parameter range. Indeed, when replacing the luminosity by the effective temperature in the exponent $q$ when re-deriving the fitting formula, the predicted \Mdot\ only changes by at most $\sim 4\%$ to those predicted by Eq. (20) in Paper II. Now, when we turn to the current grid of models we find that $q$ shows the tightest relation with \Teff, which is shown in Fig. \ref{q}. The scatter in this figure likely arises due to a combination of less significant dependencies on other parameters and the limited number of metallicity points used in our model grid.
    Finally then, we construct a relation for the mass-loss rate from a multi-linear fit including the $Z_{\ast}(\Teff)$ dependence as follows
    \begin{align}\label{recipe}
        \log \Mdot =& - 5.52 \nonumber\\
                    & + 2.39\log\left(\frac{L_{\ast}}{10^6\Lsol}\right) \nonumber\\
                    & - 1.48\log\left(\frac{M_{\rm eff}}{45\Msol}\right) \\
                    & + 2.12\log\left(\frac{\Teff}{45000 K}\right) \nonumber\\
                    & + \left(0.75-1.87\log\left(\frac{\Teff}{45000 K}\right)\right)\log\left(\frac{Z_{\ast}}{Z_{\odot}}\right) \nonumber. 
    \end{align}
    This recipe is valid within the ranges $4.5\leq\log L_{\ast}/\Lsol\leq6.0$, $15\leq M_{\ast}/\Lsol\leq80$, $15000 {\rm K}\leq\Teff\leq50000 {\rm K}$, and $0.2\leq Z_{\ast}/Z_{\odot}\leq1.0$. Despite its simple form, it is still a reasonably good representation of the results of the grid where most values of the mass-loss rates are represented by the recipe to within a factor 2. More quantitatively, taking a mean across the complete grid of the ratio between the actual calculated mass-loss rate, $\dot{M}_{\rm mod}$, and that predicted by the simple fitting formula above, $\dot{M}_{\rm fit}$, yields 1.14 (median 0.91), with almost 90 \% of all models having a ratio lying in the range 0.5 to 2. Some larger factors exist, mostly originating from the bi-stability region, where, as discussed below, an overall larger scatter on the mass-loss rate is found in the grid. This then increases also the overall scatter found between the models and our fitting formula, so that when taken across the complete grid we find a standard deviation 1.67 for the ratio $\dot{M}_{\rm mod}/\dot{M}_{\rm fit}$. By contrast, when considering only models with $T_{\rm eff} \ge 30\,000$ K, this standard deviation decreases significantly to 0.34 (consistent with Paper II). We emphasize, however, that this relatively large scatter does not affect a key conclusion found from our models with lower $T_{\rm eff}$, namely that we do not find any evidence for a systematic increase in $\dot{M}$ within the bi-stability region (see next section). As mentioned above, the models are further all computed with a standard value for the micro-turbulent velocity of 10 km/s. In Paper II we derived a dependence of \Mdot\ on \vturb, which can be applied to the predicted mass-loss rates here as well in case  \vturb\ should deviate significantly from 10 km/s. 
    
\section{Analysis} 
    
    \subsection{The bi-stability region}
        An important aspect of our new grid regards the behaviour within the so-called `bi-stability' region around $\sim$ 20 kK. Here,  
        several important metals change their ionisation stage which affects the radiative acceleration throughout the wind. One notable transition is that of Fe IV to Fe III.  
        Due to recombination of iron, \citet{Vink99} find that the mass-loss rate `jumps' by an order of magnitude when going from the hot to the cool side of the region at \Teff=25000 K \citep[some later models seem to suggest a somewhat lower \Teff=21500 K, see][]{Petrov16}.
        Recently, \citet{Krticka21} also found that \Mdot\ increased when lowering \Teff\ across the bi-stability region, at least in the high-luminosity region and with more modest factors than those obtained by Vink et al. Also these models by Krticka et al., however, are a bit different than the ones presented here, since they are based on a solution to the equation-of-motion wherein the critical point of the flow is shifted downstream from the sonic point to the supersonic parts of the flow. As such, they also find an additional dependence on `clumping' for their mass-loss rates. But as discussed in detail in Sect. 2.2, it is questionable if current techniques trying to account for clumping within 1D steady-state models of line-driven flows are really able to properly capture such effects.
        
\begin{figure*}
            \centering
            \includegraphics[width=0.49\hsize]{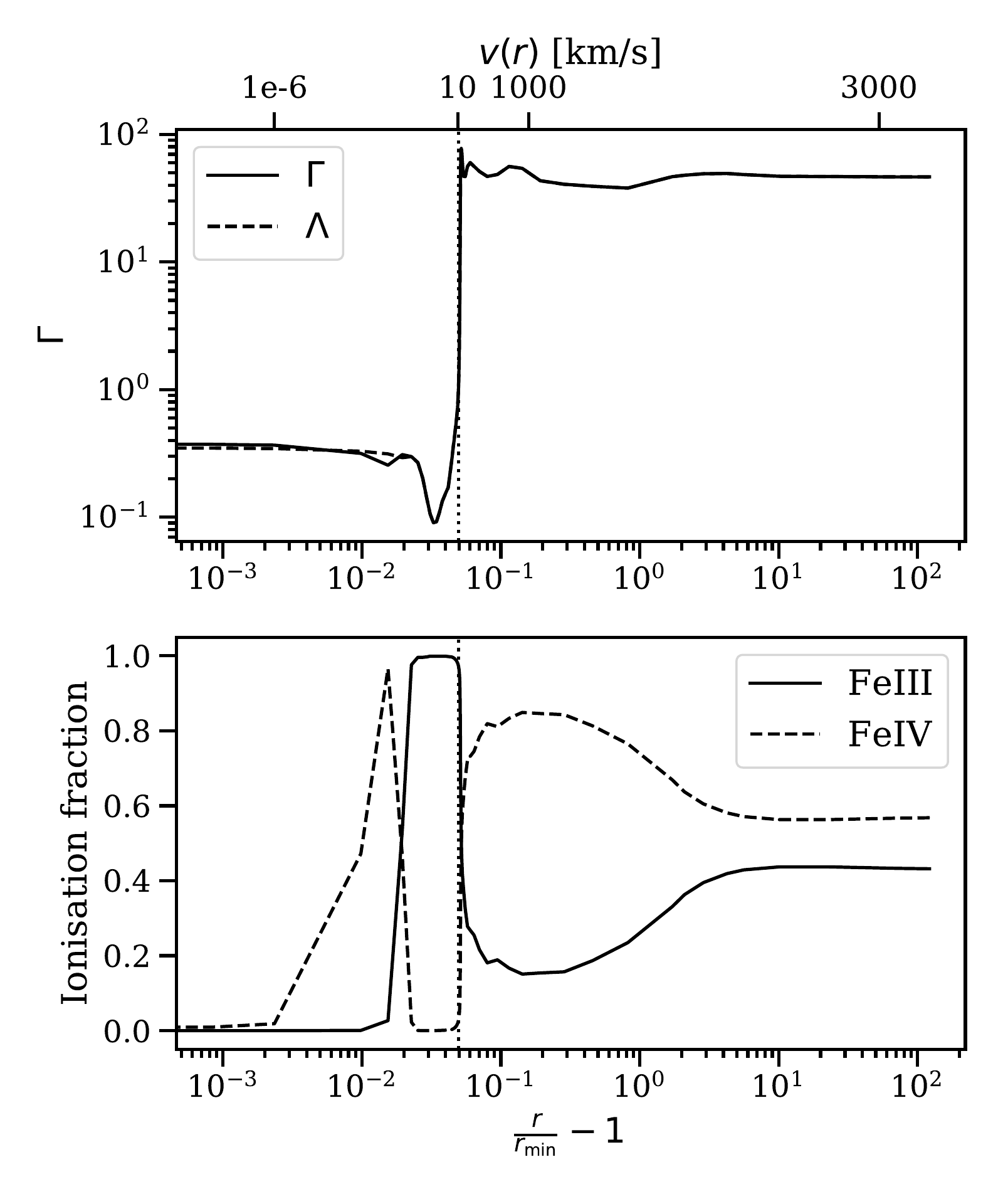}
            \includegraphics[width=0.49\hsize]{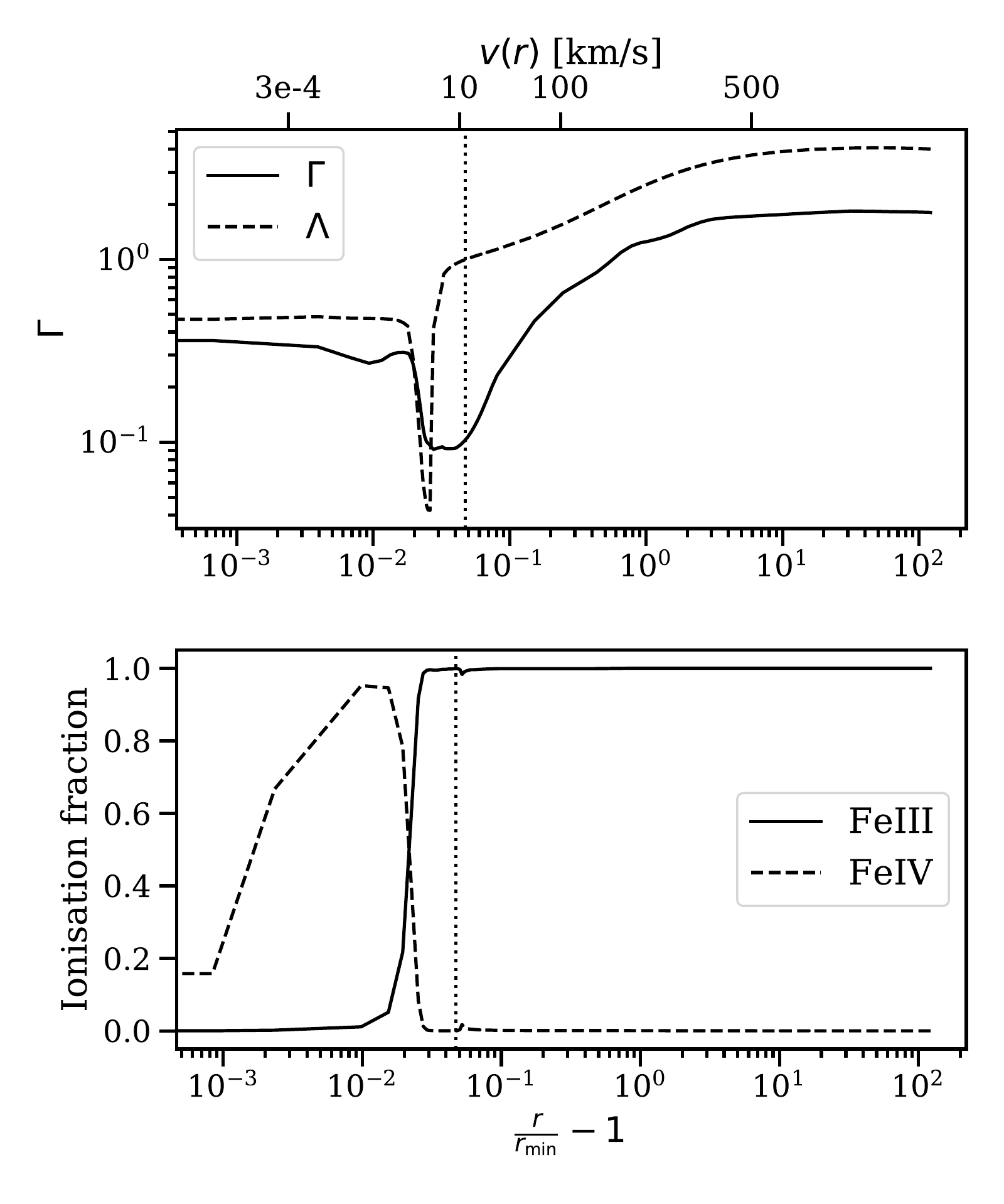}
            \caption{Wind structure for models with $\log(L_{\ast}/\Lsol) = 4.5$, $M_{\ast} = 15 \Msol$, and $\Teff=15000$K. The upper panels show $\Gamma$ and the other terms in the eom throughout the wind. The lower panels show the ionisation fractions for FeIII and FeIV. The panels on the upper left and lower left are the results for the self-consistent models (with $\Mdot=2.64\cdot10^{-10}$ \Msol/yr and $\vinf=3028$ km/s), while the panels on the upper right and lower right are the results using a $\beta$-velocity law  of $\beta = 1.0$ (with $\Mdot=4.67\cdot10^{-8}$ \Msol/yr and $\vinf=588$ km/s). The abscissa show a radius-coordinate below and the corresponding velocity on top.} 
            \label{L45_bistab}
        \end{figure*}

        In our models, we do not find an increase of \Mdot\ when crossing to the cool side of the bi-stability region.
        To further investigate this absence of a bi-stability jump, we compute additional models at 15000 K (i.e. below the previously predicted bi-stability region). We select two models from the grid, one with high and one with low luminosity, calculating the CMF-transfer by fixing the velocity-structure to a $\beta$-law and selecting $\Mdot$ as predicted by Vink et al. together with the recommended $\vinf=1.3\vesc$ and a $\beta=1.0$ (we have also tested both a lower and higher $\beta$, 0.6 and 3, with essentially the same result). Fig. \ref{L45_bistab} shows the resulting force-balance and iron ionisation balance for the low-luminosity model, including as a reference the results for the self-consistent and fully converged model with the same stellar parameters. The mass-loss rate predicted by the Vink et al. recipe is here a factor 170 higher than the resulting self-consistent \Mdot\ of the converged model. For completeness, the corresponding results for the high-luminosity models are shown in Fig. \ref{L575_bistab}, where $\Mdot_{\rm Vink}/\Mdot = 70$.
        
        The first thing to notice from these figures is that the calculated $\Gamma$ in the models using a $\beta$-velocity law does not match the other terms in the eom (i.e. the model is not hydrodynamically consistent). Particularly important is the discrepancy around the sonic point $a$, where $\Gamma(\varv=a) \approx 0.1$ in the $\beta$-law model (compared to the required $\Gamma=0.97$ for a consistent model). This points to the significantly reduced radiation force in this critical region (see also discussions in Paper I and II), which makes it impossible to drive an outflow of such a high mass loss through the sonic point using the steady-state modelling technique applied here. These results are qualitatively the same for the high-luminosity model; in this case $\Gamma = 0.46$ at the sonic point for the model with the high \Mdot\ predicted by Vink et al. 
        
        For these tests, it is clear that our detailed CMF-calculations do not provide enough radiation force to drive a flow with such a high mass-loss rate through the sonic point. As discussed in Paper I and II (see also Owocki \& Puls 1999), the basic reason for this might be related to the dip in radiative acceleration in atmospheric layers leading up to the sonic point. The `force-dip' directly arises from source-function gradients in the resonance zone which are accounted for in the CMF line-transfer. In order for the model to be dynamically consistent, this force-dip induces a very steep acceleration in near-sonic atmospheric regions and requires the mass-loss rate to be substantially reduced as compared to the $\beta$-law models displayed in Fig. 3.
        
        Another important conclusion we can draw from the figures concerns the ionisation stage of iron. Although our code does not allow us to explicitly analyse the contributions to $g_{\rm rad}$ from individual elements, the alternative models by \citet{Vink00} and \citet{Krticka21} both point to the importance of iron for determining the overall scale of the total line force in this region. The models with very high \Mdot\ and a $\beta$-velocity-law generally have higher fractions of FeIII than the consistent models in the outer part of the wind. However, when inspecting the conditions around the sonic point, we can see that both models have iron mostly in FeIII. This means that even though the consistent models have access to the lower ionisation stages of iron, there is still not enough acceleration to launch a wind with a very high mass-loss rate. To compensate for this lack of force, the model then self-adjust towards a lower mass-loss rate in order to become dynamically consistent. This then suggests that the drastic \Mdot\ increase found in earlier models in this region might simply be an artefact of not being dynamically consistent around the sonic point, and not allowing properly for the feedback between radiative and velocity accelerations. 
        The models of \citet{Vink21b}, using the set-up of \citet{Muller08}, do have dynamic solutions. However, due to their parametrised way of calculating \grad\ (essentially Eq. 14 in \citealt{Muller08}) and deriving the mass-loss rate (using a global energy constraint), these models still do not capture the dynamical properties around the sonic point stemming from the CMF-force here.
        
        The radiative acceleration predicted by \FW\ has been thoroughly benchmarked to the alternative NLTE radiative transfer code  \textsc{\large cmfgen} \citep{Hillier98}, however so far mainly focusing on the O-star and early B-star regime \citep{Puls20}. The results of the models with an effective temperature around and below the bi-stability region, have thus not yet been subjected to such an extensive calibration. The question then arises whether our atomic data at these lower temperatures include sufficient lines to represent a physically realistic result for the mass-loss rates, or if additional lines are required, which might then increase the radiative driving and result in higher mass-loss rates. To address this question, we calculated additional models from 15 kK to 22.5 kK both using \FW\ with a beta-velocity law\footnote{Since \textsc{\large cmfgen} does not solve the hydrodynamical equation we make use of $\beta$-law models for the purpose of a one to one comparison. This should produce almost identical wind and density structures, except for a subtle difference in the definition of the $\beta$-law and how the connection photosphere-wind is performed in each code.} and \textsc{\large cmfgen} (which uses a line list completely independent from that used in \FW) using exactly the same stellar and wind parameters. The results for our additional models show overall a very good agreement for the radiative acceleration between both codes, especially below and around the sonic point (which is the crucial region setting the mass-loss rate). A detailed description of these models and comparisons is given in Appendix \ref{check_grad}.
        


\subsection{Comparison to other results}

\begin{figure}
        \centering
        \includegraphics[width=\hsize]{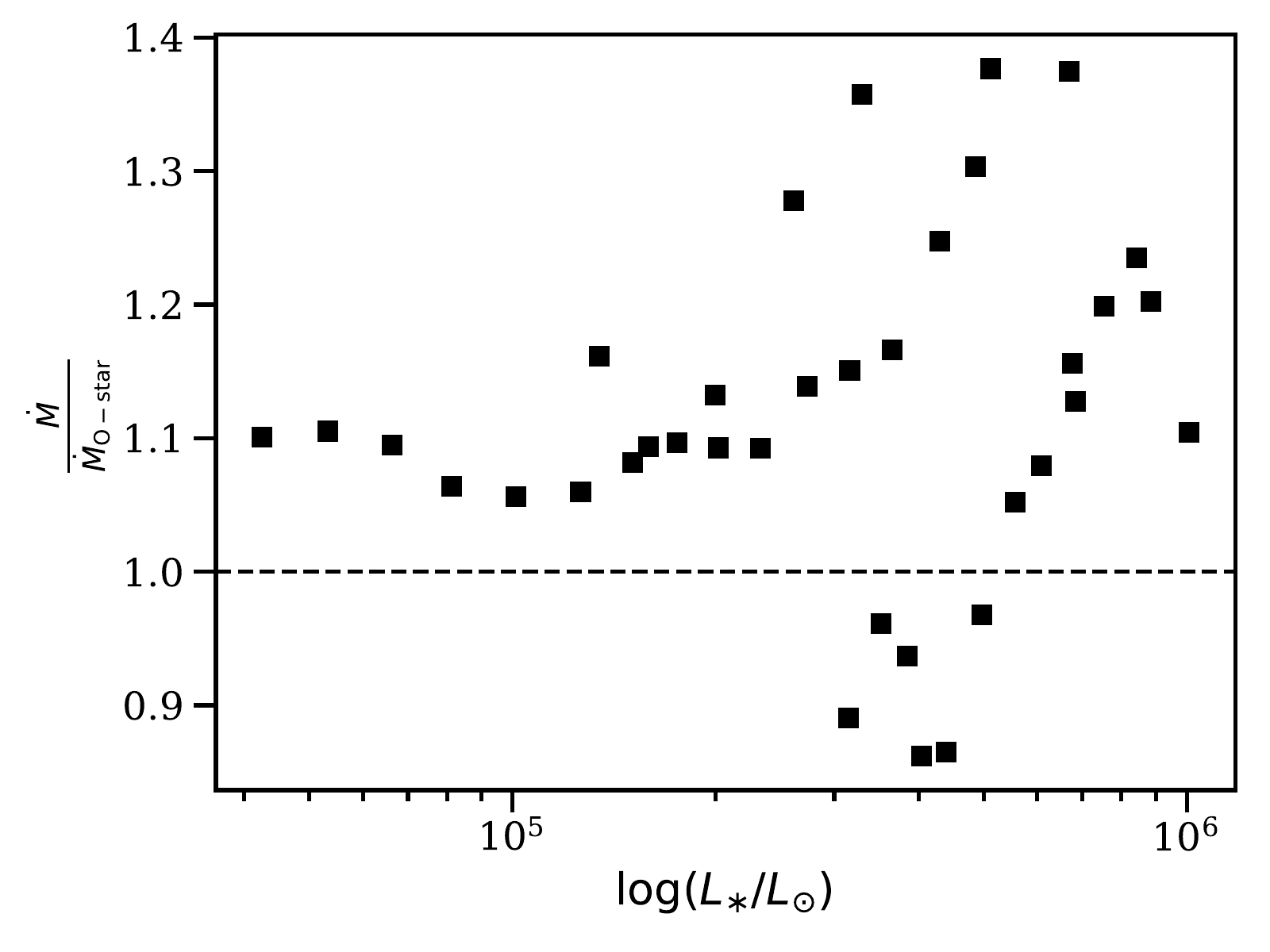}
        \caption{Ratio of \Mdot\ as computed from Eq. \eqref{recipe} and from Paper II for the O-stars presented in that paper which show a mean ratio of 1.12 and a standard deviation of 0.13.}
        \label{oldvsnew}
    \end{figure}

    Here we analyse how the mass-loss prescription from Eq. \eqref{recipe} compares to other theoretical and observational results. First, we show in Fig. \ref{oldvsnew} the predictions from equation \eqref{recipe} to those of Eq. (20) in Paper II for the set of O-stars presented in that paper (for a Galactic metallicity).
    The predictions agree well with each other, with the mean ratio only changing by 12\%, due to slightly higher \Mdot\ values in the new recipe, displaying a standard deviation of 13\%. When including O-stars of LMC and SMC metallicity, the mean changes to 1.15 and the standard deviation to 0.20. The new recipe also matches slightly better with the calculated mass-loss rates of the O-star models, because it of course captures the additional (small) effects from the mass and \Teff\ variation also within this region.
    
    \subsubsection{Comparison to alternative theoretical mass-loss recipes} 
    
\begin{figure}
        \centering
        \includegraphics[width=\hsize]{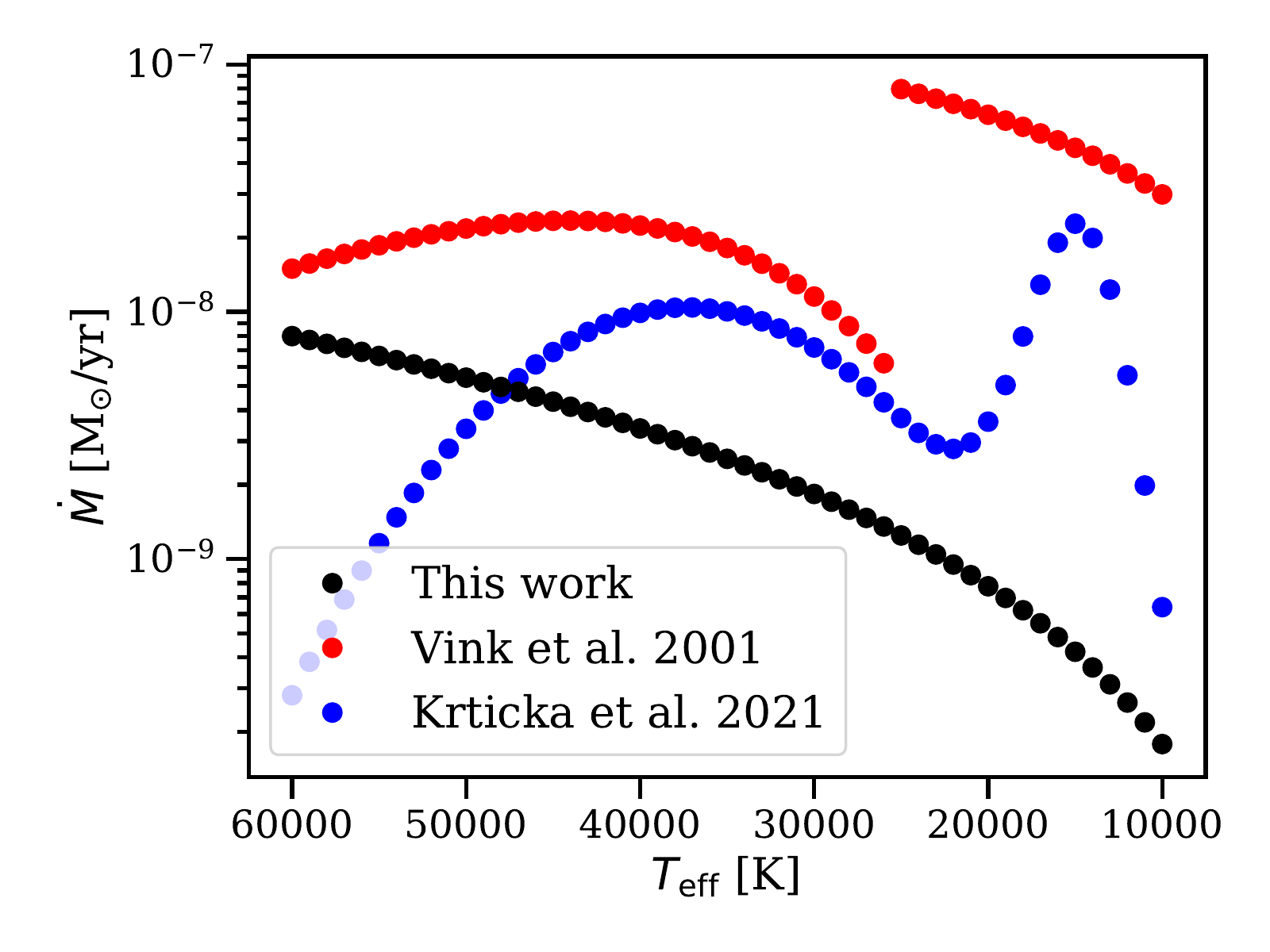}
        \includegraphics[width=\hsize]{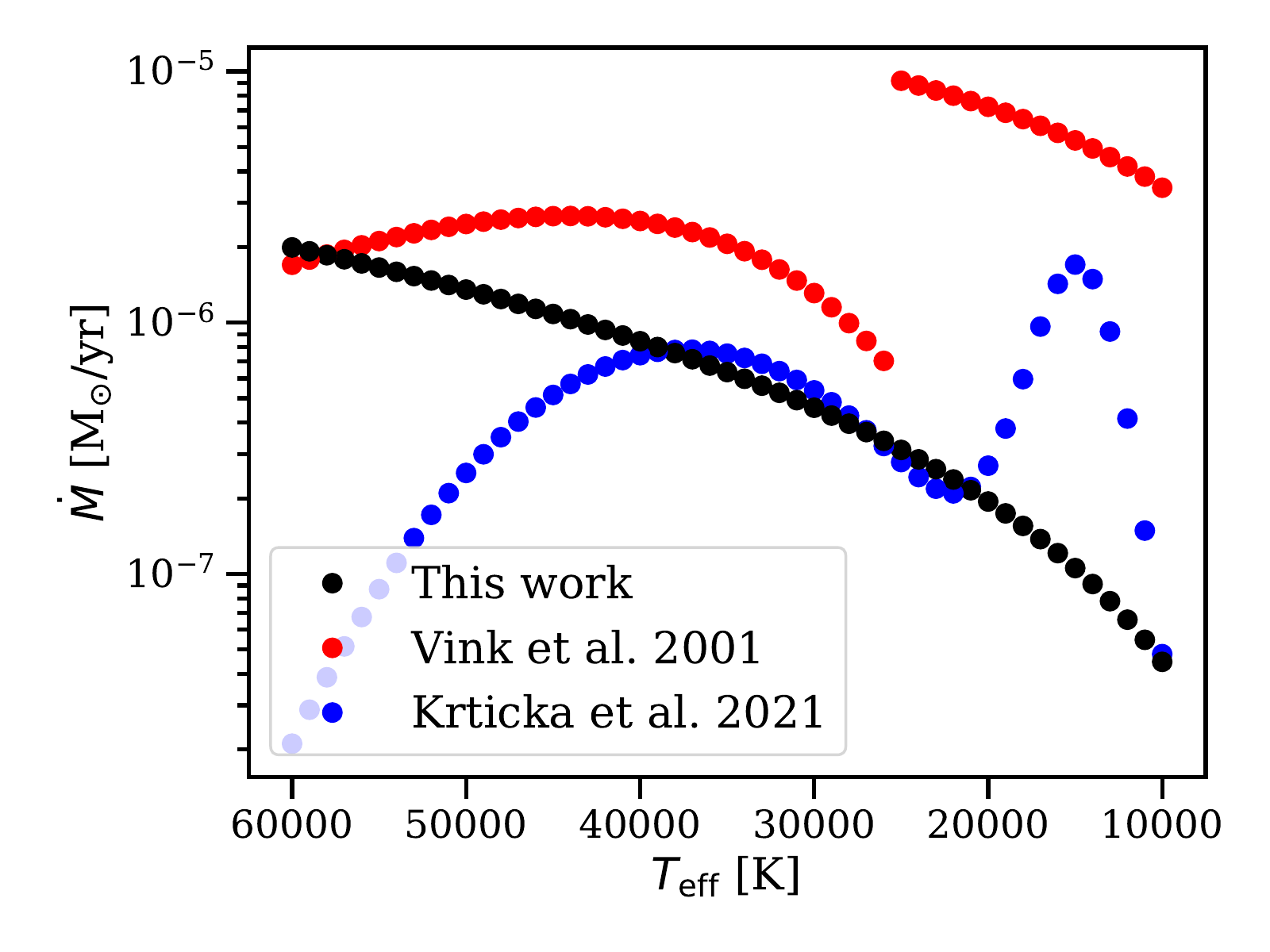}
        \caption{Mass-loss rates predicted from Eq. \eqref{recipe} (black), from the Vink et al. recipe as implemented in {\sc MESA} (red), and from the Krticka et al. recipe (blue), at constant $\log(L_{\ast}/\Lsol) = 4.5$ and $M_{\ast} = 15 \Msol$ for the upper panel and $\log(L_{\ast}/\Lsol) = 5.75$ and $M_{\ast} = 50 \Msol$ for the lower panel.}
        \label{LvsV}
    \end{figure}

    Next we compare our new prescription to other current mass-loss predictions for hot, massive stars. Fig. \ref{LvsV} shows this comparison, using the recipe from \citet{Vink00,Vink01} as implemented in the standard `Dutch scheme' in \MESA, as well as from \citet{Krticka21}, for two sequences of effective temperature at a fixed high and low luminosity, respectively.
    
    The figures show a clear reduction of the mass-loss rate as compared to the Vink et al. recipe, with slightly more pronounced differences for the low luminosity sequence. The difference-factor is about 2 to 3 on the hot side of the bi-stability region, but gets smaller when going towards higher luminosities and temperatures; for example for a hot star with \Teff>50 kK and $L_{\ast}/L_{\odot}=5.75$ the rates are in good agreement (see lower panel of Fig. \ref{LvsV}). On the cool side of the bi-stability region the difference-factor becomes on average 50, mainly because we do not predict the strong bi-stability jump that is present in this implementation of the Vink et al recipe.
    
    In the region around 30 to 40 kK, there is reasonable agreement with the \citet{Krticka21} prescription for the more luminous star, but less so when inspecting the less luminous star. This is in accordance with the results from Paper II, which show a shallower dependence of \Mdot\ on stellar luminosity for the results by \citet{Krticka18}\footnote{The results from \citet{Krticka18} for O-stars have been incorporated in the more general recipe in \citet{Krticka21} for O and B stars.} as compared to ours. There is also disagreement between prescriptions at lower and higher effective temperatures. At higher effective temperatures, the Krticka et al. recipe is actually out of the domain where their mass-loss rates are derived and a direct comparison with our rates therefore becomes difficult in this regime. At lower effective temperatures, their mass loss shows a gradual increase followed by a rather steep decline. While their increase in mass-loss rate in this region is smaller than that predicted by Vink et al., the occurrence of a local maximum still means that there is a quite large disagreement with our predictions. Because of the different modelling techniques (see discussions above), it is again difficult to judge exactly why these differences arise, but this should be further investigated in future work.
    
    The evolution of O and B type stars can vary significantly depending on whether or not this jump is included in the mass-loss recipe. \citet{Keszthelyi17} investigate its effect on the early evolution of these stars and their rotation, because the stellar winds not only remove mass but also angular momentum from the surface layers \citep{Langer98}. This angular-momentum loss in turn has an effect on the evolution through altered internal mixing and transport of angular momentum \citep{Maeder00}. \citet{Keszthelyi17} find that, in order to explain the rotational velocities observed in late B supergiants, the used rates (from \citealt{Vink00,Vink01}) have to be reduced, either by reducing the early O-star rates or by removing the occurrence of the bi-stability jump. Both reductions could also occur simultaneously if an additional source of angular momentum loss is present or if the initial rotational velocities of O-type stars that is generally assumed \citep[$\sim 300$ km/s, e.g.][]{Howarth97} are too high \citep{Simon-Diaz14}.
    
    
    \subsubsection{Comparison to observations} 
    
    In Paper II we presented a comparison of the theoretical predictions to a selected sample of observations for O-stars, where we find good overall agreement. Considering the minor deviations of these O-star predictions with the new predictions of Eq. \eqref{recipe}, as shown above, we still recover the same overall agreement with these observations. We can additionally look at the recent study by \citet{Hawcroft21}, who analyse a sample of Galactic O supergiants by means of a multi-diagnostics (UV+optical) genetic algorithm fitting accounting fully for potentially optically thick clumps and `velocity-porosity' in the line formation process (Sundqvist \& Puls 2018). These empirical results again agree to within about 10\% with our theoretical predictions.
    
    With our new, more extended grid and recipe we can now also compare to empirical mass-loss rates at cooler temperatures, particularly investigating observational evidence pointing to the occurrence/absence of a bi-stability jump. In this respect, we first compare to the maximum mass-loss rates derived from radio observations such as in \citet{Benaglia07} and \cite{RubioDiez21} who also include an analysis of Herschel/PACS continuum emission data.
    As these mass-loss rates are obtained from radio-diagnostics assuming an unclumped outermost wind, the derived results are upper limits ($\Mdot_{\rm max}$). Any inclusion of clumping would reduce the observationally inferred mass-loss rate by a factor of $\sqrt{f_{cl}}$, where $f_{cl}=\frac{\langle\rho^2\rangle}{\langle\rho\rangle^2}$ is the clumping factor in the radio-emitting region. While \citet{Benaglia07} find a modestly increased wind efficiency ($\Mdot\vinf/L_{\ast} c$) around \Teff=21.5 kK, both studies still find a continuation of the decrease of mass-loss rates below the bi-stability jump (i.e. they do not find any evidence for a jump in mass loss across the bi-stability region), in agreement with the overall trend of this paper. Actually, these upper-limit observed rates are even significantly lower than the rates predicted by the Vink et al. recipe. The measured upper limits are, however, also consistently above our predictions for all stars. The inclusion of clumping in the empirical mass-loss rates would then reduce the values to numbers comparable to our predictions, assuming a clumping factor of about 5-20 in the radio-emitting region. This shift could of course be altered if the clumping factor should depend on spectral type, which then would shift the different observations by different factors \citep[see e.g.][]{Driessen19}, which has been discussed in \citet{Markova08}.  
    
    A comparison can also be made with analysis of stellar and wind parameters derived from the optical. \citet{Crowther06} and \citet{Markova08} derive mass-loss rates from the optical spectra of B-type supergiants using \textsc{\large cmfgen} \citep{Hillier98} and \FW, respectively. \citet{Crowther06} find no significant increase in the mass-loss rate below $\Teff=24000$K even without the inclusion of clumping (which would further decrease their empirical results). Similarly \citet{Markova08} find that the observed mass-loss rates below the bi-stability region either decreases or increases only marginally. All observational results above thus suggest that the previously predicted bi-stability jump in mass-loss is not present (at least not for stars of not too high luminosity), but instead that rates in this region seem to follow similar scalings as for O-stars (see also \citealt{Keszthelyi17}). 
    
    We do however predict high terminal wind speeds which are not generally found in the observational literature. Across the grid, we have mean values of about 3300 km/s and $\vinf/\varv_{\rm esc, eff} \approx 4.5$ (where $\vinf/\varv_{\rm esc, eff} = \sqrt{\frac{2G M_{\ast}(1-\Gamma_e)}{R_{\ast}}}$), which corresponds to a value of $\vinf/\vesc \approx 4$ when using the regular escape speed, not reduced by electron scattering.
    This value does not significantly decrease for lower effective temperatures.
    These overall high terminal wind speeds may point towards either missing physics in the outer wind-regions of the models or inaccurate empirical terminal velocity estimates when the sharp edge of the ultraviolet P-Cygni lines is not clearly present in the observed spectrum. \citet{Lagae21} computed time-dependent, line-driven instability (LDI) hydrodynamic simulations of weak-wind stars, and find that a significant portion of the weak wind is shock-heated and unable to cool down efficiently, so that the UV-line opacity becomes significantly reduced. This might then not only lead to erroneous empirical mass-loss estimates, but also to underestimated terminal speeds derived from UV-line profile fitting (see their Fig. 5 for an illustration).
    Additionally, as studied in Paper II, a shock-heated outer wind could also reduce the radiation force in those regions, such that if we were to account for the shock heated gas the predicted \vinf\ might be substantially reduced. On the other hand, significant portions of high-temperature gas is not really expected for winds of higher density (which can cool efficiently by radiative losses), and thus it is questionable whether 
    this effect could also reduce terminal wind speeds in the high-luminosity parts of the HRD. Further investigations are clearly needed here, both regarding obtaining better empirical constraints on terminal wind speeds and in terms of theoretical analysis of additional physics not included in steady-state models. 

\subsection{Stellar evolution}
    Our recipe for \Mdot\ can readily be implemented into evolution codes. In this paper, we chose to work with \MESA\ \citep{Paxton19}. \MESA, as mentioned above, is a modelling tool for the structure and evolution of astrophysical stellar objects and allows for simple adaptations of its input physics modules. The \texttt{other\_wind} routine offers the option to calculate and apply a mass-loss rate from the stellar parameters at a certain time step. Here we implemented the mass-loss recipe from Sect. \ref{sec:recipe} and applied it to all hot, hydrogen-rich stars. As mentioned before, the wind models are computed for a fixed $Y_{\rm He} = 0.1$. As the star evolves, this value can change when, for example, material is mixed and brought to the surface. This effect is not incorporated in the mass-loss recipe, as to keep the amount of variable parameters of the grid reasonably small, and it is still unclear how much it would affect the mass-loss predictions.
    
    To analyse the impact of our new predicted mass-loss on the evolution of stars with a mass between 15 and 80 \Msol, we perform a differential study comparing to previous models. Naturally, many physical parameters can influence the evolution of the massive star, for example mixing processes due to overshooting from a convective region or from rotation. The purpose of the differential study here, however, is to isolate the effect of mass loss. As a baseline for the evolution models, we chose to work with a similar setup as utilised in MIST \citep[MESA Isochrones \& Stellar Tracks][]{Dotter16,Choi16}. Their stellar tracks are also computed using \MESA\ and cover a wide range in mass and metallicity. \citet{Choi16} present comparisons both with observational constraints and already existing models from literature. This allows them to compile a set of prescriptions and parameters applied throughout the various evolution stages (see their Table 1); in Tab. \ref{table:MESA} we mention some of these that are particularly relevant for our massive-star evolution models.
    \begin{table}
        \caption{Parameters and prescriptions of a massive-star model in \MESA. See Tab. 1 in \citet{Choi16} for an extended list of the prescriptions.}
        \centering
        \begin{tabular}{l l l}
        \hline\hline
        Ingredient & Adopted physics & Details \\
        \hline
        Mass loss & This paper & $\Teff>11$kK, $X_{\rm surf}>0.4$ \\
          & de Jager et al. & $\Teff<10$kK \\
          & Nugis \& Lamers & $\Teff>10$kK, $X_{\rm surf}<0.4$ \\
         Overshoot & $f_{\rm ov, core} = 0.016 $ &  exponential overshoot\\
          & $f_{\rm ov, envelope} = 0.0174 $ & \\
         Convection & Ledoux + MLT & Includes superadiabatic \\
          & & convection (MLT++)\\
         Rotation & $\varv_{\rm init} = 350$ km/s & $\Omega/\Omega_{\rm crit}=0.4$ \\
        \hline
        \end{tabular}
        \label{table:MESA}
    \end{table}
    Concerning the mass loss, we apply three different prescriptions depending on temperature and evolution phase. For hot ($\Teff>11$kK), hydrogen-rich phases we apply our mass-loss formula \eqref{recipe}. 
    This implementation is different to previous stellar evolution calculation, such as those from MIST, where instead the mass-loss rates from \citet{Vink00,Vink01} are applied. When the star evolves to the cooler side of the HR-diagram ($\Teff<10$kK), we switch to the empirical rates by \citet{deJager88}. Even though our grid only extends to $\Teff=15$kK, we still apply our mass-loss formula until 10 kK, because an extrapolation of these rates to these cooler temperatures probably yields more realistic results than extrapolating the de Jager rates, which are primarily used to describe the winds of cool supergiants. For the hot ($\Teff>10$kK), stripped (with a surface hydrogen mass fraction $X_{\rm surf}<0.4$) stars we apply the WR-star mass loss from \citet{Nugis00}. In the transition regions between two prescriptions, a linear interpolation between those recipes is used. This transition occurs between 10 kK and 11 kK between the cool and hot-star recipes and between $X_{\rm surf} = 0.4$ and a few percent below for this surface abundance to transition to the stripped-star recipe.
    
    Additionally, the table mentions parameters concerning internal transport of energy and matter. For example, exponential overshoot is applied extending the reach of the convective core and any convective layers in the envelope. This convection is described by the conventional mixing length theory \citep[MLT,][]{Bohm-Vitense58} and the location of convection is determined by the Ledoux criterion which compares the temperature gradient with the adiabatic and chemical gradients. Furthermore, to treat the radiation-dominated envelopes of massive stars, where standard MLT is not always applicable, convection is treated by artificially reducing the superadiabaticity, effectively assuming energy is carried away by some other means than radiation or standard convection \citep[MLT++,][]{Paxton13}. A last prescription we mention is rotation, where the star has an initial angular rotation of $\Omega = 0.4 \Omega_{\rm crit}$, with $\Omega_{\rm crit}$ the critical rotation where the centrifugal force equals gravity.
    
\begin{figure}
        \centering
        \includegraphics[width=\hsize]{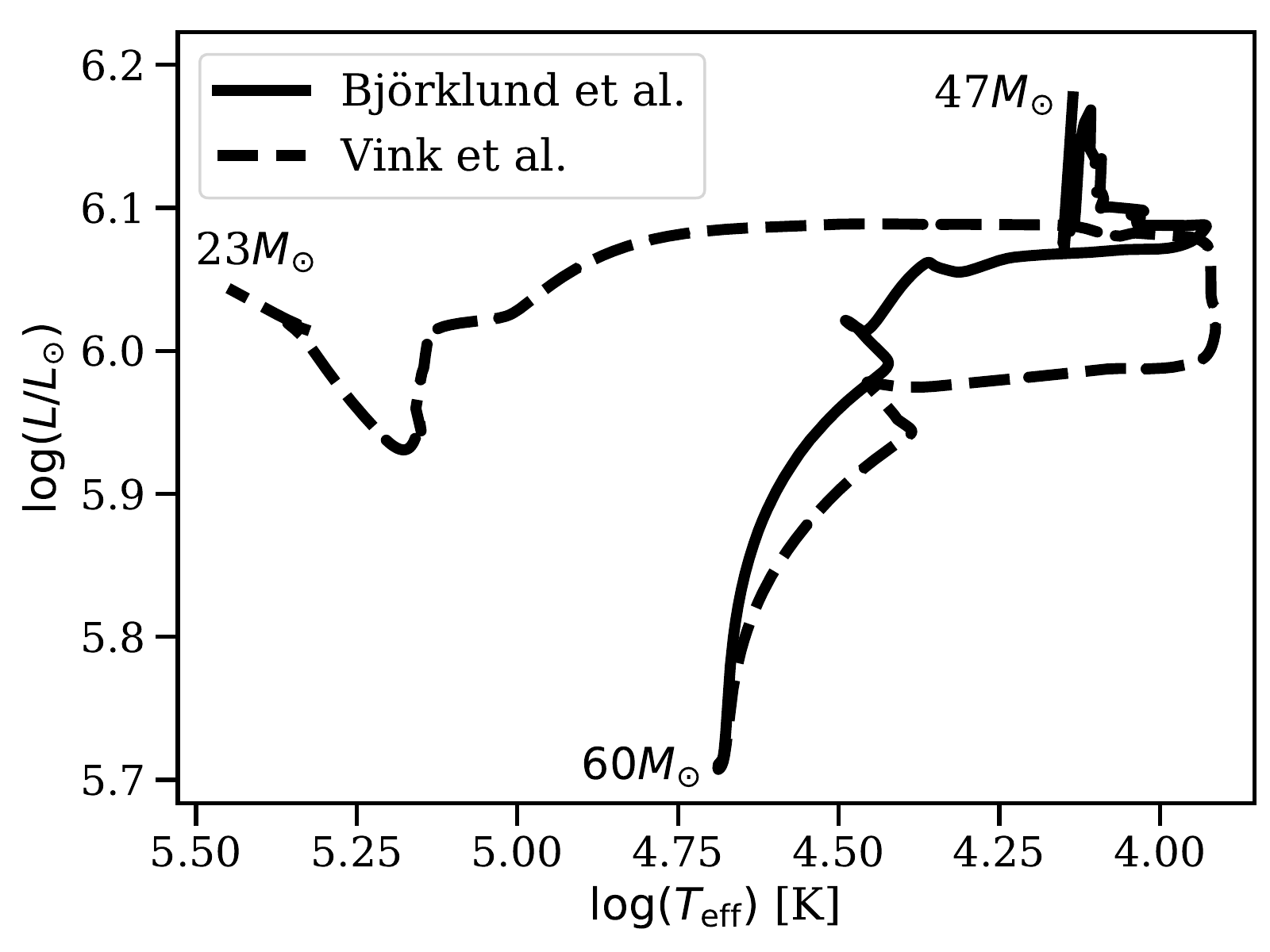}
        \caption{HR-diagram for a star with an initial mass of $M_{\ast}=60$ \Msol and an initial surface rotation of 350 km/s. The two tracks are calculated assuming different mass-loss prescriptions.}
        \label{M60}
    \end{figure}

    We have used this setup to calculate tracks of stars with different initial masses, comparing to models using an equivalent setup however applying the (previously standard) Vink et al. recipe instead. Fig. \ref{M60} shows the resulting  Hertzprung-Russel (HR) diagram for a star with initial mass $M_{\ast}=60$ \Msol\ and $Z_{\ast} = Z_{\odot}$, and here below we now discuss two important aspects that are seen in these stellar tracks resulting from our new mass-loss rates. 
    
    \subsubsection{No WR stars from standard line-driven wind stripping?}
        Classical Wolf-Rayet (WR) stars are created when the hydrogen rich envelope of the massive star is somehow stripped, 
        which in evolution models typically occurs through wind-stripping or binary interactions. In the single-star models considered in this paper, only the first mechanism can be active.  
        
        
        In Fig. \ref{M60} we illustrate the evolution of a star with an initial mass 60 \Msol\ and solar metallicity. It can be directly seen that when applying our new rates the core-helium burning phase ends without the star ever evolving into a hot WR star. On the other hand, when applying the previous higher rates, mass loss through line-driven wind-stripping is indeed enough to make the star evolve back to the blue parts of the HRD. 
        The typical differences in mass loss between these two characteristic models are factors of $\approx 3$ on the main sequence and $\approx 30$ for the cooler regions (where the previous rates experience the bi-stability jump, see discussion in previous section). The model where lower mass loss rates are applied do not evolve to become a WR star because the outer envelope is not stripped before helium-burning ends after which the evolutionary time-scale becomes too short to expel a lot more mass. This happens because mass is lost more slowly than when adopting the previous rates while the lifetime of the star is even shorter, as the star will live as a higher-mass star (compare for example 54 \Msol\ versus 45 \Msol\ in the B supergiant stage after the main sequence). So the combined effect of a shorter lifetime and reduced mass loss makes it a lot more difficult for a star to evolve to the WR phase. For the 60 \Msol\ star, this conclusion remains unchanged, also when changing the condition when the Nugis \& Lamers rates are applied. When in this model, these mass-loss rates are applied from $X_{\rm surf}<0.6$ instead of 0.4, more mass is lost (about 8 \Msol\ difference at the end of the evolution), but not enough to strip the star.
        
        This differential comparison thus shows that with the implementation of our new rates in massive-star models, the formation of classical WR-stars through standard, steady-state line-driven wind-stripping becomes difficult, even at Galactic metalicities. Indeed, from further test-models using the same set-up, we find that we must increase the initial mass to $>100 \Msol$ in order to form an evolved, hot WR helium star with our new mass-loss rates\footnote{These models applied an extrapolation of the mass-loss rates beyond the grid at high mass and luminosity, where as discussed above our prescription might not fully capture the nature of these stars' winds.}.
        Because such very massive stars are also very rare, this suggests that the formation of classical WR-stars might be dominated by other processes.
        
        As mentioned above, one such alternative formation channel for classical WR stars regards envelope-stripping via binary interaction. However, observational campaigns do show signs of single WR-stars, even in low-metallicity environments such as the SMC \citep{Shenar16,Shenar20}, so that there may a need for yet another channel by which stars strip their envelopes. One such possibility regards variable mass loss from high-luminosity LBV-like stars that exceed the Eddington limit already in deep sub-surface layers \citep[see e.g.][]{Owocki15}. Indeed, such stars are observed, but the evolution model does not pass through such a phase. As discussed in Sect. 3, LBV stars are not included in our current model grid, both because of technical difficulties with modelling deep-seated wind initiation and since the core-assumption of a steady-state outflow becomes highly questionable in this regime. As such, our recipe may well underestimate the mass loss in this regime\footnote{even though some first 3D radiation-hydrodynamical simulations of LBV envelopes by Jiang et al. (2018) only found moderate mass-loss rates, not significantly higher than those predicted by our steady-state recipe here.}.

    \subsubsection{Massive black holes}
        The final fate of many massive stars is to die in a supernova (SN) explosion, leaving behind a neutron star or a black hole. The final mass of the black hole is largely determined by the core mass of the progenitor before the SN. Burning stages beyond carbon-burning are unstable for stars that have helium cores more massive that 30 \Msol\ \citep{Woosley07}. The creation of an electron-positron pair requires energy at high temperatures, which reduces the amount of energy and momentum available to provide a pressure balance counteracting gravity. Contraction at high temperatures then becomes unstable, which is called the pair instability (PI). As stars get more massive, the mass of the helium core increases and results in a more violent implosion as a result of this PI. This is believed to cause stars with a He-core mass greater than 64 \Msol\ to become PI supernova where a single pulse disrupts the entire star, leaving no remnant \citep{Glatzel85,Heger02,Ober83,Bond84}. The implosion typically happens in a series of pulses when nuclear burning is not sufficient to counter the PI (this mechanism is called pulsational pair-instability, PPI, and is believed to give rise to PPI supernovae \citealt{Woosley17}). During the pulses, the core contracts, initiates nuclear burning, and then expands and cools. The cycle repeats itself until the mass of the Helium core is sufficiently lowered to avoid the PPI. Due to this, the final helium core masses that manage to avoid PPI, or that have come down as a result of PPI falls in the range $35-45 \Msol\ $ \citep{Woosley17}.
        
        The mass of the remnant black hole is then dependent on both the He-core mass (to determine whether a PI or PPI may occur) and the mass of the remaining hydrogen envelope that gets added to the final black hole mass. In the case of the weaker winds that we predict in this paper, massive stars retain part of their hydrogen envelopes also at the end stages of their lives. Fig. \ref{M60} shows the final mass after helium burning for a 60 \Msol\ Galactic star. We assume this, similarly as in \citet{Belczynski20}, to be a good representation of the final mass before core collapse. As directly seen in the figure, using our new rates we find that the total final mass is more than double than that predicted by the comparison model employing the previous standard mass-loss recipe. 
        For the template model displayed here we find this final mass to be between 45-50 \Msol\, but for even higher initial masses (if the star can avoid being disrupted) this can be even higher. 
        Indeed, \citet{Belczynski20} found final masses as high as 70 \Msol\ from their evolution calculations using an ad hoc (factor of 5) reduction of the previously standard mass-loss rates. Here we thus recover qualitatively the same result, however with the key difference that our result does not depend on any ad-hoc reduction of mass-loss rates, but rather is a natural consequence of our new theoretical predictions. 
        We note again though, that the massive stars that would create such massive black holes indeed tend to evolve towards the LBV-like regime in the HRD, where (see discussion above) our current steady-state recipe might still not be sufficient to properly describe the mass loss behaviour.  
        
        When considering a lower overall mass-loss, either by reducing the initial mass of the star (to below about 30 \Msol) or by reducing the metallicity, the difference between using different mass-loss prescriptions in the prediction of the black-hole mass becomes less important. \citet{Vink21a} show, for example, that with the use of the Vink et al. rates, high-mass black holes are created at a metallicity of $Z_{\ast}\leq0.1Z_{\odot}$ from stars with an initial mass of 90..100 \Msol. However, if the metallicity is not as low, the difference in the mass-loss recipe will affect how much of the envelope is blown away and thus how much mass is available to collapse into the black hole.

\section{Summary}
    We have computed a large grid of atmosphere and wind models for hot and massive stars. The grid covers a large parameter range in stellar luminosity, mass, and effective temperature for the Galaxy and the Large and Small Magellanic Clouds. 
    From the dynamically-consistent, steady-state wind models we obtain theoretical mass-loss rates and investigate their dependencies on fundamental stellar parameters. The results are captured by the mass-loss prescription of Eq. \eqref{recipe}. As in the previous O-star calculations by \citet{Bjorklund21}, we find strong trends of the mass-loss rate with $L_{\ast}$ and $Z_{\ast}$, which are supplemented here with additional dependencies on $M_{\rm eff}$ and \Teff\ in order to properly capture the behaviour throughout our larger grid. We additionally found that we could best represent the results of this grid when including also a temperature dependence inside the exponent for the metallicity.
    
    We obtain generally good agreements between our predictions and observations of O-stars in the Galaxy (for studies where clumping is taken into account when deriving empirical mass-loss rates). Similarly, when comparing to observations using radio-diagnostics we find that empirical upper limits of OB-star mass-loss rates are higher than our predictions with factors around 3.5, corresponding to reasonable clumping factors of 12.
    
    When comparing the new recipe to existing predictions of \citet{Vink00,Vink01}, we again find a downwards shift by approximately a factor of 3 for O-stars \citep{Bjorklund21}, with somewhat smaller values for the most luminous stars. However, more notably, a much larger discrepancy is found for stars below 25 kK. Here we do not find any evidence of a bi-stability jump, which increases the mass-loss rate by an order of magnitude or more according to the mass-loss recipe by \citet{Vink00}. Additional models where we try to drive a very high mass-loss rate using a fixed $\beta$-velocity law (instead of iterating towards a consistent structure), show that there is not enough radiation force to drive such a large amount of mass through the wind sonic point. 
    The absence of a significant mass-loss increase in the bi-stability region also agrees with empirical studies of H$\alpha$ \citep{Markova08,Crowther06} and radio emission \citep{Benaglia07,RubioDiez21}. 
    
    Evolution models of massive stars using the new predictions show significant differences as compared to equivalent models employing the previous standard mass-loss rates. Here, we discuss two main influences of such decreased mass-loss rates in stellar evolution: first, with the lower mass-loss rates single stars have difficulties expelling their hydrogen-rich envelopes, which thus hinders them to turn into evolved, classical Wolf-Rayet stars at the end of their lives. Second, since the mass of the black hole remnant depends on the final mass of its progenitor, the lower mass-loss rates potentially allow for the creation of high-mass black holes even at higher metallicities. One key uncertainty with (the mass loss within) such massive-star evolution models, however, is that currently there is no adequate treatment of the potentially significant mass-loss from variable LBV-like stages \citep{Smith14}, which might ultimately help the massive star to strip its hydrogen-envelope and still create a WR star despite the lower steady-state mass-loss rates found here. 

\begin{acknowledgements}

RB and JOS acknowledge support from the Odysseus program of the Belgian Research Foundation Flanders (FWO) under grant G0H9218N. We thank the referee for their constructive and useful comments on the manuscript. 
      
\end{acknowledgements}

%
%

\bibliographystyle{aa} 
\bibliography{all_papers.bib} 

\begin{thebibliography}{76}
\expandafter\ifx\csname natexlab\endcsname\relax\def\natexlab#1{#1}\fi

\bibitem[{{Abbott} {et~al.}(2016{\natexlab{a}}){Abbott}, {Abbott}, {Abbott},
  {Abernathy}, {Acernese}, {Ackley}, {Adams}, {Adams}, {Addesso}, {Adhikari},
  {Adya}, {Affeldt}, {Agathos}, {Agatsuma}, {Aggarwal}, {Aguiar}, {Aiello},
  {Ain}, {Ajith}, {Allen}, {Allocca}, {Altin}, {Anderson}, {Anderson}, {Arai},
  {Araya}, {Arceneaux}, {Areeda}, {Arnaud}, {Arun}, {Ascenzi}, {Ashton}, {Ast},
  {Aston}, {Astone}, {Aufmuth}, {Aulbert}, {Babak}, {Bacon}, {Bader}, {Baker},
  {Baldaccini}, {Ballardin}, {Ballmer}, {Barayoga}, {Barclay}, {Barish},
  {Barker}, {Barone}, {Barr}, {Barsotti}, {Barsuglia}, {Barta}, {Bartlett},
  {Bartos}, {Bassiri}, {Basti}, {Batch}, {Baune}, {Bavigadda}, {Bazzan},
  {Behnke}, {Bejger}, {Belczynski}, {Bell}, {Bell}, {Berger}, {Bergman},
  {Bergmann}, {Berry}, {Bersanetti}, {Bertolini}, {Betzwieser}, {Bhagwat},
  {Bhandare}, {Bilenko}, {Billingsley}, {Birch}, {Birney}, {Biscans}, {Bisht},
  {Bitossi}, {Biwer}, {Bizouard}, {Blackburn}, {Blair}, {Blair}, {Blair},
  {Bloemen}, {Bock}, {Bodiya}, {Boer}, {Bogaert}, {Bogan}, {Bohe}, {Bojtos},
  {Bond}, {Bondu}, {Bonnand}, {Boom}, {Bork}, {Boschi}, {Bose}, {Bouffanais},
  {Bozzi}, {Bradaschia}, {Brady}, {Braginsky}, {Branchesi}, {Brau}, {Briant},
  {Brillet}, {Brinkmann}, {Brisson}, {Brockill}, {Brooks}, {Brown}, {Brown},
  {Brown}, {Buchanan}, {Buikema}, {Bulik}, {Bulten}, {Buonanno}, {Buskulic},
  {Buy}, {Byer}, {Cadonati}, {Cagnoli}, {Cahillane}, {Calder{\'o}n Bustillo},
  {Callister}, {Calloni}, {Camp}, {Cannon}, {Cao}, {Capano}, {Capocasa},
  {Carbognani}, {Caride}, {Casanueva Diaz}, {Casentini}, {Caudill},
  {Cavagli{\`a}}, {Cavalier}, {Cavalieri}, {Cella}, {Cepeda}, {Cerboni
  Baiardi}, {Cerretani}, {Cesarini}, {Chakraborty}, {Chalermsongsak},
  {Chamberlin}, {Chan}, {Chao}, {Charlton}, {Chassande-Mottin}, {Chen}, {Chen},
  {Cheng}, {Chincarini}, {Chiummo}, {Cho}, {Cho}, {Chow}, {Christensen}, {Chu},
  {Chua}, {Chung}, {Ciani}, {Clara}, {Clark}, {Cleva}, {Coccia}, {Cohadon},
  {Colla}, {Collette}, {Cominsky}, {Constancio}, {Conte}, {Conti}, {Cook},
  {Corbitt}, {Cornish}, {Corsi}, {Cortese}, {Costa}, {Coughlin}, {Coughlin},
  {Coulon}, {Countryman}, {Couvares}, {Cowan}, {Coward}, {Cowart}, {Coyne},
  {Coyne}, {Craig}, {Creighton}, {Cripe}, {Crowder}, {Cumming}, {Cunningham},
  {Cuoco}, {Dal Canton}, {Danilishin}, {D'Antonio}, {Danzmann}, {Darman},
  {Dattilo}, {Dave}, {Daveloza}, {Davier}, {Davies}, {Daw}, {Day}, {DeBra},
  {Debreczeni}, {Degallaix}, {De Laurentis}, {Del{\'e}glise}, {Del Pozzo},
  {Denker}, {Dent}, {Dereli}, {Dergachev}, {DeRosa}, {DeRosa}, {DeSalvo},
  {Dhurandhar}, {D{\'\i}az}, {Di Fiore}, {Di Giovanni}, {Di Lieto}, {Di Pace},
  {Di Palma}, {Di Virgilio}, {Dojcinoski}, {Dolique}, {Donovan}, {Dooley},
  {Doravari}, {Douglas}, {Downes}, {Drago}, {Drever}, {Driggers}, {Du},
  {Ducrot}, {Dwyer}, {Edo}, {Edwards}, {Effler}, {Eggenstein}, {Ehrens},
  {Eichholz}, {Eikenberry}, {Engels}, {Essick}, {Etzel}, {Evans}, {Evans},
  {Everett}, {Factourovich}, {Fafone}, {Fair}, {Fairhurst}, {Fan}, {Fang},
  {Farinon}, {Farr}, {Farr}, {Favata}, {Fays}, {Fehrmann}, {Fejer}, {Ferrante},
  {Ferreira}, {Ferrini}, {Fidecaro}, {Fiori}, {Fiorucci}, {Fisher}, {Flaminio},
  {Fletcher}, {Fournier}, {Franco}, {Frasca}, {Frasconi}, {Frei}, {Freise},
  {Frey}, {Frey}, {Fricke}, {Fritschel}, {Frolov}, {Fulda}, {Fyffe}, {Gabbard},
  {Gair}, {Gammaitoni}, {Gaonkar}, {Garufi}, {Gatto}, {Gaur}, {Gehrels},
  {Gemme}, {Gendre}, {Genin}, {Gennai}, {George}, {Gergely}, {Germain},
  {Ghosh}, {Ghosh}, {Giaime}, {Giardina}, {Giazotto}, {Gill}, {Glaefke},
  {Goetz}, {Goetz}, {Gondan}, {Gonz{\'a}lez}, {Gonzalez Castro}, {Gopakumar},
  {Gordon}, {Gorodetsky}, {Gossan}, {Gosselin}, {Gouaty}, {Graef}, {Graff},
  {Granata}, {Grant}, {Gras}, {Gray}, {Greco}, {Green}, {Groot}, {Grote},
  {Grunewald}, {Guidi}, {Guo}, {Gupta}, {Gupta}, {Gushwa}, {Gustafson},
  {Gustafson}, {Hacker}, {Hall}, {Hall}, {Hammond}, {Haney}, {Hanke}, {Hanks},
  {Hanna}, {Hannam}, {Hanson}, {Hardwick}, {Harms}, {Harry}, {Harry}, {Hart},
  {Hartman}, {Haster}, {Haughian}, {Heidmann}, {Heintze}, {Heitmann}, {Hello},
  {Hemming}, {Hendry}, {Heng}, {Hennig}, {Heptonstall}, {Heurs}, {Hild},
  {Hoak}, {Hodge}, {Hofman}, {Hollitt}, {Holt}, {Holz}, {Hopkins}, {Hosken},
  {Hough}, {Houston}, {Howell}, {Hu}, {Huang}, {Huerta}, {Huet}, {Hughey},
  {Husa}, {Huttner}, {Huynh-Dinh}, {Idrisy}, {Indik}, {Ingram}, {Inta}, {Isa},
  {Isac}, {Isi}, {Islas}, {Isogai}, {Iyer}, {Izumi}, {Jacqmin}, {Jang}, {Jani},
  {Jaranowski}, {Jawahar}, {Jim{\'e}nez-Forteza}, {Johnson}, {Jones}, {Jones},
  {Jonker}, {Ju}, {K}, {Kalaghatgi}, {Kalogera}, {Kandhasamy}, {Kang},
  {Kanner}, {Karki}, {Kasprzack}, {Katsavounidis}, {Katzman}, {Kaufer}, {Kaur},
  {Kawabe}, {Kawazoe}, {K{\'e}f{\'e}lian}, {Kehl}, {Keitel}, {Kelley}, {Kells},
  {Kennedy}, {Key}, {Khalaidovski}, {Khalili}, {Khan}, {Khan}, {Khan},
  {Khazanov}, {Kijbunchoo}, {Kim}, {Kim}, {Kim}, {Kim}, {Kim}, {Kim}, {King},
  {King}, {Kinzel}, {Kissel}, {Kleybolte}, {Klimenko}, {Koehlenbeck},
  {Kokeyama}, {Koley}, {Kondrashov}, {Kontos}, {Korobko}, {Korth}, {Kowalska},
  {Kozak}, {Kringel}, {Krishnan}, {Kr{\'o}lak}, {Krueger}, {Kuehn}, {Kumar},
  {Kuo}, {Kutynia}, {Lackey}, {Landry}, {Lange}, {Lantz}, {Lasky}, {Lazzarini},
  {Lazzaro}, {Leaci}, {Leavey}, {Lebigot}, {Lee}, {Lee}, {Lee}, {Lee}, {Lenon},
  {Leonardi}, {Leong}, {Leroy}, {Letendre}, {Levin}, {Levine}, {Li}, {Libson},
  {Littenberg}, {Lockerbie}, {Logue}, {Lombardi}, {Lord}, {Lorenzini},
  {Loriette}, {Lormand}, {Losurdo}, {Lough}, {L{\"u}ck}, {Lundgren}, {Luo},
  {Lynch}, {Ma}, {MacDonald}, {Machenschalk}, {MacInnis}, {Macleod},
  {Maga{\~n}a-Sandoval}, {Magee}, {Mageswaran}, {Majorana}, {Maksimovic},
  {Malvezzi}, {Man}, {Mandel}, {Mandic}, {Mangano}, {Mansell}, {Manske},
  {Mantovani}, {Marchesoni}, {Marion}, {M{\'a}rka}, {M{\'a}rka}, {Markosyan},
  {Maros}, {Martelli}, {Martellini}, {Martin}, {Martin}, {Martynov}, {Marx},
  {Mason}, {Masserot}, {Massinger}, {Masso-Reid}, {Matichard}, {Matone},
  {Mavalvala}, {Mazumder}, {Mazzolo}, {McCarthy}, {McClelland}, {McCormick},
  {McGuire}, {McIntyre}, {McIver}, {McManus}, {McWilliams}, {Meacher},
  {Meadors}, {Meidam}, {Melatos}, {Mendell}, {Mendoza-Gandara}, {Mercer},
  {Merilh}, {Merzougui}, {Meshkov}, {Messenger}, {Messick}, {Meyers},
  {Mezzani}, {Miao}, {Michel}, {Middleton}, {Mikhailov}, {Milano}, {Miller},
  {Millhouse}, {Minenkov}, {Ming}, {Mirshekari}, {Mishra}, {Mitra},
  {Mitrofanov}, {Mitselmakher}, {Mittleman}, {Moggi}, {Mohan}, {Mohapatra},
  {Montani}, {Moore}, {Moore}, {Moraru}, {Moreno}, {Morriss}, {Mossavi},
  {Mours}, {Mow-Lowry}, {Mueller}, {Mueller}, {Muir}, {Mukherjee}, {Mukherjee},
  {Mukherjee}, {Mukund}, {Mullavey}, {Munch}, {Murphy}, {Murray}, {Mytidis},
  {Nardecchia}, {Naticchioni}, {Nayak}, {Necula}, {Nedkova}, {Nelemans},
  {Neri}, {Neunzert}, {Newton}, {Nguyen}, {Nielsen}, {Nissanke}, {Nitz},
  {Nocera}, {Nolting}, {Normandin}, {Nuttall}, {Oberling}, {Ochsner}, {O'Dell},
  {Oelker}, {Ogin}, {Oh}, {Oh}, {Ohme}, {Oliver}, {Oppermann}, {Oram},
  {O'Reilly}, {O'Shaughnessy}, {Ottaway}, {Ottens}, {Overmier}, {Owen}, {Pai},
  {Pai}, {Palamos}, {Palashov}, {Palomba}, {Pal-Singh}, {Pan}, {Pankow},
  {Pannarale}, {Pant}, {Paoletti}, {Paoli}, {Papa}, {Paris}, {Parker},
  {Pascucci}, {Pasqualetti}, {Passaquieti}, {Passuello}, {Patricelli},
  {Patrick}, {Pearlstone}, {Pedraza}, {Pedurand}, {Pekowsky}, {Pele}, {Penn},
  {Perreca}, {Phelps}, {Piccinni}, {Pichot}, {Piergiovanni}, {Pierro},
  {Pillant}, {Pinard}, {Pinto}, {Pitkin}, {Poggiani}, {Popolizio}, {Post},
  {Powell}, {Prasad}, {Predoi}, {Premachandra}, {Prestegard}, {Price},
  {Prijatelj}, {Principe}, {Privitera}, {Prix}, {Prodi}, {Prokhorov},
  {Puncken}, {Punturo}, {Puppo}, {P{\"u}rrer}, {Qi}, {Qin}, {Quetschke},
  {Quintero}, {Quitzow-James}, {Raab}, {Rabeling}, {Radkins}, {Raffai}, {Raja},
  {Rakhmanov}, {Rapagnani}, {Raymond}, {Razzano}, {Re}, {Read}, {Reed},
  {Regimbau}, {Rei}, {Reid}, {Reitze}, {Rew}, {Reyes}, {Ricci}, {Riles},
  {Robertson}, {Robie}, {Robinet}, {Rocchi}, {Rolland}, {Rollins}, {Roma},
  {Romano}, {Romano}, {Romanov}, {Romie}, {Rosi{\'n}ska}, {Rowan},
  {R{\"u}diger}, {Ruggi}, {Ryan}, {Sachdev}, {Sadecki}, {Sadeghian}, {Salconi},
  {Saleem}, {Salemi}, {Samajdar}, {Sammut}, {Sanchez}, {Sandberg}, {Sandeen},
  {Sanders}, {Sassolas}, {Sathyaprakash}, {Saulson}, {Sauter}, {Savage},
  {Sawadsky}, {Schale}, {Schilling}, {Schmidt}, {Schmidt}, {Schnabel},
  {Schofield}, {Sch{\"o}nbeck}, {Schreiber}, {Schuette}, {Schutz}, {Scott},
  {Scott}, {Sellers}, {Sentenac}, {Sequino}, {Sergeev}, {Serna}, {Setyawati},
  {Sevigny}, {Shaddock}, {Shah}, {Shahriar}, {Shaltev}, {Shao}, {Shapiro},
  {Shawhan}, {Sheperd}, {Shoemaker}, {Shoemaker}, {Siellez}, {Siemens}, {Sigg},
  {Silva}, {Simakov}, {Singer}, {Singer}, {Singh}, {Singh}, {Singhal},
  {Sintes}, {Slagmolen}, {Smith}, {Smith}, {Smith}, {Son}, {Sorazu},
  {Sorrentino}, {Souradeep}, {Srivastava}, {Staley}, {Steinke}, {Steinlechner},
  {Steinlechner}, {Steinmeyer}, {Stephens}, {Stevenson}, {Stone}, {Strain},
  {Straniero}, {Stratta}, {Strauss}, {Strigin}, {Sturani}, {Stuver},
  {Summerscales}, {Sun}, {Sutton}, {Swinkels}, {Szczepa{\'n}czyk}, {Tacca},
  {Talukder}, {Tanner}, {T{\'a}pai}, {Tarabrin}, {Taracchini}, {Taylor},
  {Theeg}, {Thirugnanasambandam}, {Thomas}, {Thomas}, {Thomas}, {Thorne},
  {Thorne}, {Thrane}, {Tiwari}, {Tiwari}, {Tokmakov}, {Tomlinson}, {Tonelli},
  {Torres}, {Torrie}, {T{\"o}yr{\"a}}, {Travasso}, {Traylor}, {Trifir{\`o}},
  {Tringali}, {Trozzo}, {Tse}, {Turconi}, {Tuyenbayev}, {Ugolini},
  {Unnikrishnan}, {Urban}, {Usman}, {Vahlbruch}, {Vajente}, {Valdes}, {van
  Bakel}, {van Beuzekom}, {van den Brand}, {van den Broeck}, {Vander-Hyde},
  {van der Schaaf}, {van Heijningen}, {van Veggel}, {Vardaro}, {Vass},
  {Vas{\'u}th}, {Vaulin}, {Vecchio}, {Vedovato}, {Veitch}, {Veitch},
  {Venkateswara}, {Verkindt}, {Vetrano}, {Vicer{\'e}}, {Vinciguerra}, {Vine},
  {Vinet}, {Vitale}, {Vo}, {Vocca}, {Vorvick}, {Voss}, {Vousden}, {Vyatchanin},
  {Wade}, {Wade}, {Wade}, {Walker}, {Wallace}, {Walsh}, {Wang}, {Wang}, {Wang},
  {Wang}, {Wang}, {Ward}, {Warner}, {Was}, {Weaver}, {Wei}, {Weinert},
  {Weinstein}, {Weiss}, {Welborn}, {Wen}, {We{\ss}els}, {Westphal}, {Wette},
  {Whelan}, {White}, {Whiting}, {Williams}, {Williamson}, {Willis}, {Willke},
  {Wimmer}, {Winkler}, {Wipf}, {Wittel}, {Woan}, {Worden}, {Wright}, {Wu},
  {Yablon}, {Yam}, {Yamamoto}, {Yancey}, {Yap}, {Yu}, {Yvert}, {Zadro{\.z}ny},
  {Zangrando}, {Zanolin}, {Zendri}, {Zevin}, {Zhang}, {Zhang}, {Zhang},
  {Zhang}, {Zhao}, {Zhou}, {Zhou}, {Zhu}, {Zucker}, {Zuraw}, {and}, {Zweizig},
  {LIGO Scientific Collaboration}, \& {Virgo Collaboration}}]{Abbott16}
{Abbott}, B.~P., {Abbott}, R., {Abbott}, T.~D., {et~al.} 2016{\natexlab{a}},
  \apjl, 818, L22

\bibitem[{{Abbott} {et~al.}(2016{\natexlab{b}}){Abbott}, {Abbott}, {Abbott},
  {Abernathy}, {Acernese}, {Ackley}, {Adams}, {Adams}, {Addesso}, {Adhikari},
  {Adya}, {Affeldt}, {Agathos}, {Agatsuma}, {Aggarwal}, {Aguiar}, {Aiello},
  {Ain}, {Ajith}, {Allen}, {Allocca}, {Altin}, {Anderson}, {Anderson}, {Arai},
  {Araya}, {Arceneaux}, {Areeda}, {Arnaud}, {Arun}, {Ascenzi}, {Ashton}, {Ast},
  {Aston}, {Astone}, {Aufmuth}, {Aulbert}, {Babak}, {Bacon}, {Bader}, {Baker},
  {Baldaccini}, {Ballardin}, {Ballmer}, {Barayoga}, {Barclay}, {Barish},
  {Barker}, {Barone}, {Barr}, {Barsotti}, {Barsuglia}, {Barta}, {Bartlett},
  {Bartos}, {Bassiri}, {Basti}, {Batch}, {Baune}, {Bavigadda}, {Bazzan},
  {Behnke}, {Bejger}, {Belczynski}, {Bell}, {Bell}, {Berger}, {Bergman},
  {Bergmann}, {Berry}, {Bersanetti}, {Bertolini}, {Betzwieser}, {Bhagwat},
  {Bhandare}, {Bilenko}, {Billingsley}, {Birch}, {Birney}, {Biscans}, {Bisht},
  {Bitossi}, {Biwer}, {Bizouard}, {Blackburn}, {Blair}, {Blair}, {Blair},
  {Bloemen}, {Bock}, {Bodiya}, {Boer}, {Bogaert}, {Bogan}, {Bohe}, {Bojtos},
  {Bond}, {Bondu}, {Bonnand}, {Boom}, {Bork}, {Boschi}, {Bose}, {Bouffanais},
  {Bozzi}, {Bradaschia}, {Brady}, {Braginsky}, {Branchesi}, {Brau}, {Briant},
  {Brillet}, {Brinkmann}, {Brisson}, {Brockill}, {Brooks}, {Brown}, {Brown},
  {Brown}, {Buchanan}, {Buikema}, {Bulik}, {Bulten}, {Buonanno}, {Buskulic},
  {Buy}, {Byer}, {Cadonati}, {Cagnoli}, {Cahillane}, {Calder{\'o}n Bustillo},
  {Callister}, {Calloni}, {Camp}, {Cannon}, {Cao}, {Capano}, {Capocasa},
  {Carbognani}, {Caride}, {Casanueva Diaz}, {Casentini}, {Caudill},
  {Cavagli{\`a}}, {Cavalier}, {Cavalieri}, {Cella}, {Cepeda}, {Cerboni
  Baiardi}, {Cerretani}, {Cesarini}, {Chakraborty}, {Chalermsongsak},
  {Chamberlin}, {Chan}, {Chao}, {Charlton}, {Chassande-Mottin}, {Chen}, {Chen},
  {Cheng}, {Chincarini}, {Chiummo}, {Cho}, {Cho}, {Chow}, {Christensen}, {Chu},
  {Chua}, {Chung}, {Ciani}, {Clara}, {Clark}, {Cleva}, {Coccia}, {Cohadon},
  {Colla}, {Collette}, {Cominsky}, {Constancio}, {Conte}, {Conti}, {Cook},
  {Corbitt}, {Cornish}, {Corsi}, {Cortese}, {Costa}, {Coughlin}, {Coughlin},
  {Coulon}, {Countryman}, {Couvares}, {Cowan}, {Coward}, {Cowart}, {Coyne},
  {Coyne}, {Craig}, {Creighton}, {Cripe}, {Crowder}, {Cumming}, {Cunningham},
  {Cuoco}, {Dal Canton}, {Danilishin}, {D'Antonio}, {Danzmann}, {Darman},
  {Dattilo}, {Dave}, {Daveloza}, {Davier}, {Davies}, {Daw}, {Day}, {DeBra},
  {Debreczeni}, {Degallaix}, {De Laurentis}, {Del{\'e}glise}, {Del Pozzo},
  {Denker}, {Dent}, {Dereli}, {Dergachev}, {DeRosa}, {DeRosa}, {DeSalvo},
  {Dhurandhar}, {D{\'\i}az}, {Di Fiore}, {Di Giovanni}, {Di Lieto}, {Di Pace},
  {Di Palma}, {Di Virgilio}, {Dojcinoski}, {Dolique}, {Donovan}, {Dooley},
  {Doravari}, {Douglas}, {Downes}, {Drago}, {Drever}, {Driggers}, {Du},
  {Ducrot}, {Dwyer}, {Edo}, {Edwards}, {Effler}, {Eggenstein}, {Ehrens},
  {Eichholz}, {Eikenberry}, {Engels}, {Essick}, {Etzel}, {Evans}, {Evans},
  {Everett}, {Factourovich}, {Fafone}, {Fair}, {Fairhurst}, {Fan}, {Fang},
  {Farinon}, {Farr}, {Farr}, {Favata}, {Fays}, {Fehrmann}, {Fejer}, {Ferrante},
  {Ferreira}, {Ferrini}, {Fidecaro}, {Fiori}, {Fiorucci}, {Fisher}, {Flaminio},
  {Fletcher}, {Fournier}, {Franco}, {Frasca}, {Frasconi}, {Frei}, {Freise},
  {Frey}, {Frey}, {Fricke}, {Fritschel}, {Frolov}, {Fulda}, {Fyffe}, {Gabbard},
  {Gair}, {Gammaitoni}, {Gaonkar}, {Garufi}, {Gatto}, {Gaur}, {Gehrels},
  {Gemme}, {Gendre}, {Genin}, {Gennai}, {George}, {Gergely}, {Germain},
  {Ghosh}, {Ghosh}, {Giaime}, {Giardina}, {Giazotto}, {Gill}, {Glaefke},
  {Goetz}, {Goetz}, {Gondan}, {Gonz{\'a}lez}, {Gonzalez Castro}, {Gopakumar},
  {Gordon}, {Gorodetsky}, {Gossan}, {Gosselin}, {Gouaty}, {Graef}, {Graff},
  {Granata}, {Grant}, {Gras}, {Gray}, {Greco}, {Green}, {Groot}, {Grote},
  {Grunewald}, {Guidi}, {Guo}, {Gupta}, {Gupta}, {Gushwa}, {Gustafson},
  {Gustafson}, {Hacker}, {Hall}, {Hall}, {Hammond}, {Haney}, {Hanke}, {Hanks},
  {Hanna}, {Hannam}, {Hanson}, {Hardwick}, {Harms}, {Harry}, {Harry}, {Hart},
  {Hartman}, {Haster}, {Haughian}, {Heidmann}, {Heintze}, {Heitmann}, {Hello},
  {Hemming}, {Hendry}, {Heng}, {Hennig}, {Heptonstall}, {Heurs}, {Hild},
  {Hoak}, {Hodge}, {Hofman}, {Hollitt}, {Holt}, {Holz}, {Hopkins}, {Hosken},
  {Hough}, {Houston}, {Howell}, {Hu}, {Huang}, {Huerta}, {Huet}, {Hughey},
  {Husa}, {Huttner}, {Huynh-Dinh}, {Idrisy}, {Indik}, {Ingram}, {Inta}, {Isa},
  {Isac}, {Isi}, {Islas}, {Isogai}, {Iyer}, {Izumi}, {Jacqmin}, {Jang}, {Jani},
  {Jaranowski}, {Jawahar}, {Jim{\'e}nez-Forteza}, {Johnson}, {Jones}, {Jones},
  {Jonker}, {Ju}, {K}, {Kalaghatgi}, {Kalogera}, {Kandhasamy}, {Kang},
  {Kanner}, {Karki}, {Kasprzack}, {Katsavounidis}, {Katzman}, {Kaufer}, {Kaur},
  {Kawabe}, {Kawazoe}, {K{\'e}f{\'e}lian}, {Kehl}, {Keitel}, {Kelley}, {Kells},
  {Kennedy}, {Key}, {Khalaidovski}, {Khalili}, {Khan}, {Khan}, {Khan},
  {Khazanov}, {Kijbunchoo}, {Kim}, {Kim}, {Kim}, {Kim}, {Kim}, {Kim}, {King},
  {King}, {Kinzel}, {Kissel}, {Kleybolte}, {Klimenko}, {Koehlenbeck},
  {Kokeyama}, {Koley}, {Kondrashov}, {Kontos}, {Korobko}, {Korth}, {Kowalska},
  {Kozak}, {Kringel}, {Krishnan}, {Kr{\'o}lak}, {Krueger}, {Kuehn}, {Kumar},
  {Kuo}, {Kutynia}, {Lackey}, {Landry}, {Lange}, {Lantz}, {Lasky}, {Lazzarini},
  {Lazzaro}, {Leaci}, {Leavey}, {Lebigot}, {Lee}, {Lee}, {Lee}, {Lee}, {Lenon},
  {Leonardi}, {Leong}, {Leroy}, {Letendre}, {Levin}, {Levine}, {Li}, {Libson},
  {Littenberg}, {Lockerbie}, {Logue}, {Lombardi}, {Lord}, {Lorenzini},
  {Loriette}, {Lormand}, {Losurdo}, {Lough}, {L{\"u}ck}, {Lundgren}, {Luo},
  {Lynch}, {Ma}, {MacDonald}, {Machenschalk}, {MacInnis}, {Macleod},
  {Maga{\~n}a-Sandoval}, {Magee}, {Mageswaran}, {Majorana}, {Maksimovic},
  {Malvezzi}, {Man}, {Mandel}, {Mandic}, {Mangano}, {Mansell}, {Manske},
  {Mantovani}, {Marchesoni}, {Marion}, {M{\'a}rka}, {M{\'a}rka}, {Markosyan},
  {Maros}, {Martelli}, {Martellini}, {Martin}, {Martin}, {Martynov}, {Marx},
  {Mason}, {Masserot}, {Massinger}, {Masso-Reid}, {Matichard}, {Matone},
  {Mavalvala}, {Mazumder}, {Mazzolo}, {McCarthy}, {McClelland}, {McCormick},
  {McGuire}, {McIntyre}, {McIver}, {McManus}, {McWilliams}, {Meacher},
  {Meadors}, {Meidam}, {Melatos}, {Mendell}, {Mendoza-Gandara}, {Mercer},
  {Merilh}, {Merzougui}, {Meshkov}, {Messenger}, {Messick}, {Meyers},
  {Mezzani}, {Miao}, {Michel}, {Middleton}, {Mikhailov}, {Milano}, {Miller},
  {Millhouse}, {Minenkov}, {Ming}, {Mirshekari}, {Mishra}, {Mitra},
  {Mitrofanov}, {Mitselmakher}, {Mittleman}, {Moggi}, {Mohan}, {Mohapatra},
  {Montani}, {Moore}, {Moore}, {Moraru}, {Moreno}, {Morriss}, {Mossavi},
  {Mours}, {Mow-Lowry}, {Mueller}, {Mueller}, {Muir}, {Mukherjee}, {Mukherjee},
  {Mukherjee}, {Mukund}, {Mullavey}, {Munch}, {Murphy}, {Murray}, {Mytidis},
  {Nardecchia}, {Naticchioni}, {Nayak}, {Necula}, {Nedkova}, {Nelemans},
  {Neri}, {Neunzert}, {Newton}, {Nguyen}, {Nielsen}, {Nissanke}, {Nitz},
  {Nocera}, {Nolting}, {Normandin}, {Nuttall}, {Oberling}, {Ochsner}, {O'Dell},
  {Oelker}, {Ogin}, {Oh}, {Oh}, {Ohme}, {Oliver}, {Oppermann}, {Oram},
  {O'Reilly}, {O'Shaughnessy}, {Ottaway}, {Ottens}, {Overmier}, {Owen}, {Pai},
  {Pai}, {Palamos}, {Palashov}, {Palomba}, {Pal-Singh}, {Pan}, {Pankow},
  {Pannarale}, {Pant}, {Paoletti}, {Paoli}, {Papa}, {Paris}, {Parker},
  {Pascucci}, {Pasqualetti}, {Passaquieti}, {Passuello}, {Patricelli},
  {Patrick}, {Pearlstone}, {Pedraza}, {Pedurand}, {Pekowsky}, {Pele}, {Penn},
  {Perreca}, {Phelps}, {Piccinni}, {Pichot}, {Piergiovanni}, {Pierro},
  {Pillant}, {Pinard}, {Pinto}, {Pitkin}, {Poggiani}, {Popolizio}, {Post},
  {Powell}, {Prasad}, {Predoi}, {Premachandra}, {Prestegard}, {Price},
  {Prijatelj}, {Principe}, {Privitera}, {Prix}, {Prodi}, {Prokhorov},
  {Puncken}, {Punturo}, {Puppo}, {P{\"u}rrer}, {Qi}, {Qin}, {Quetschke},
  {Quintero}, {Quitzow-James}, {Raab}, {Rabeling}, {Radkins}, {Raffai}, {Raja},
  {Rakhmanov}, {Rapagnani}, {Raymond}, {Razzano}, {Re}, {Read}, {Reed},
  {Regimbau}, {Rei}, {Reid}, {Reitze}, {Rew}, {Reyes}, {Ricci}, {Riles},
  {Robertson}, {Robie}, {Robinet}, {Rocchi}, {Rolland}, {Rollins}, {Roma},
  {Romano}, {Romano}, {Romanov}, {Romie}, {Rosi{\'n}ska}, {Rowan},
  {R{\"u}diger}, {Ruggi}, {Ryan}, {Sachdev}, {Sadecki}, {Sadeghian}, {Salconi},
  {Saleem}, {Salemi}, {Samajdar}, {Sammut}, {Sanchez}, {Sandberg}, {Sandeen},
  {Sanders}, {Sassolas}, {Sathyaprakash}, {Saulson}, {Sauter}, {Savage},
  {Sawadsky}, {Schale}, {Schilling}, {Schmidt}, {Schmidt}, {Schnabel},
  {Schofield}, {Sch{\"o}nbeck}, {Schreiber}, {Schuette}, {Schutz}, {Scott},
  {Scott}, {Sellers}, {Sentenac}, {Sequino}, {Sergeev}, {Serna}, {Setyawati},
  {Sevigny}, {Shaddock}, {Shah}, {Shahriar}, {Shaltev}, {Shao}, {Shapiro},
  {Shawhan}, {Sheperd}, {Shoemaker}, {Shoemaker}, {Siellez}, {Siemens}, {Sigg},
  {Silva}, {Simakov}, {Singer}, {Singer}, {Singh}, {Singh}, {Singhal},
  {Sintes}, {Slagmolen}, {Smith}, {Smith}, {Smith}, {Son}, {Sorazu},
  {Sorrentino}, {Souradeep}, {Srivastava}, {Staley}, {Steinke}, {Steinlechner},
  {Steinlechner}, {Steinmeyer}, {Stephens}, {Stevenson}, {Stone}, {Strain},
  {Straniero}, {Stratta}, {Strauss}, {Strigin}, {Sturani}, {Stuver},
  {Summerscales}, {Sun}, {Sutton}, {Swinkels}, {Szczepa{\'n}czyk}, {Tacca},
  {Talukder}, {Tanner}, {T{\'a}pai}, {Tarabrin}, {Taracchini}, {Taylor},
  {Theeg}, {Thirugnanasambandam}, {Thomas}, {Thomas}, {Thomas}, {Thorne},
  {Thorne}, {Thrane}, {Tiwari}, {Tiwari}, {Tokmakov}, {Tomlinson}, {Tonelli},
  {Torres}, {Torrie}, {T{\"o}yr{\"a}}, {Travasso}, {Traylor}, {Trifir{\`o}},
  {Tringali}, {Trozzo}, {Tse}, {Turconi}, {Tuyenbayev}, {Ugolini},
  {Unnikrishnan}, {Urban}, {Usman}, {Vahlbruch}, {Vajente}, {Valdes}, {van
  Bakel}, {van Beuzekom}, {van den Brand}, {van den Broeck}, {Vander-Hyde},
  {van der Schaaf}, {van Heijningen}, {van Veggel}, {Vardaro}, {Vass},
  {Vas{\'u}th}, {Vaulin}, {Vecchio}, {Vedovato}, {Veitch}, {Veitch},
  {Venkateswara}, {Verkindt}, {Vetrano}, {Vicer{\'e}}, {Vinciguerra}, {Vine},
  {Vinet}, {Vitale}, {Vo}, {Vocca}, {Vorvick}, {Voss}, {Vousden}, {Vyatchanin},
  {Wade}, {Wade}, {Wade}, {Walker}, {Wallace}, {Walsh}, {Wang}, {Wang}, {Wang},
  {Wang}, {Wang}, {Ward}, {Warner}, {Was}, {Weaver}, {Wei}, {Weinert},
  {Weinstein}, {Weiss}, {Welborn}, {Wen}, {We{\ss}els}, {Westphal}, {Wette},
  {Whelan}, {White}, {Whiting}, {Williams}, {Williamson}, {Willis}, {Willke},
  {Wimmer}, {Winkler}, {Wipf}, {Wittel}, {Woan}, {Worden}, {Wright}, {Wu},
  {Yablon}, {Yam}, {Yamamoto}, {Yancey}, {Yap}, {Yu}, {Yvert}, {Zadro{\.z}ny},
  {Zangrando}, {Zanolin}, {Zendri}, {Zevin}, {Zhang}, {Zhang}, {Zhang},
  {Zhang}, {Zhao}, {Zhou}, {Zhou}, {Zhu}, {Zucker}, {Zuraw}, {and}, {Zweizig},
  {LIGO Scientific Collaboration}, \& {Virgo Collaboration}}]{Abbott16b}
{Abbott}, B.~P., {Abbott}, R., {Abbott}, T.~D., {et~al.} 2016{\natexlab{b}},
  \apjl, 818, L22

\bibitem[{{Asplund} {et~al.}(2009){Asplund}, {Grevesse}, {Sauval}, \&
  {Scott}}]{asplund09}
{Asplund}, M., {Grevesse}, N., {Sauval}, A.~J., \& {Scott}, P. 2009, \araa, 47,
  481

\bibitem[{{Belczynski} {et~al.}(2020){Belczynski}, {Hirschi}, {Kaiser}, {Liu},
  {Casares}, {Lu}, {O'Shaughnessy}, {Heger}, {Justham}, \&
  {Soria}}]{Belczynski20}
{Belczynski}, K., {Hirschi}, R., {Kaiser}, E.~A., {et~al.} 2020, \apj, 890, 113

\bibitem[{{Benaglia} {et~al.}(2007){Benaglia}, {Vink}, {Mart{\'\i}}, {Ma{\'\i}z
  Apell{\'a}niz}, {Koribalski}, \& {Crowther}}]{Benaglia07}
{Benaglia}, P., {Vink}, J.~S., {Mart{\'\i}}, J., {et~al.} 2007, \aap, 467, 1265

\bibitem[{{Bj{\"o}rklund} {et~al.}(2021){Bj{\"o}rklund}, {Sundqvist}, {Puls},
  \& {Najarro}}]{Bjorklund21}
{Bj{\"o}rklund}, R., {Sundqvist}, J.~O., {Puls}, J., \& {Najarro}, F. 2021,
  \aap, 648, A36

\bibitem[{{Bloecker}(1995)}]{Blocker95}
{Bloecker}, T. 1995, \aap, 297, 727

\bibitem[{{B{\"o}hm-Vitense}(1958)}]{Bohm-Vitense58}
{B{\"o}hm-Vitense}, E. 1958, \zap, 46, 108

\bibitem[{{Bond} {et~al.}(1984){Bond}, {Arnett}, \& {Carr}}]{Bond84}
{Bond}, J.~R., {Arnett}, W.~D., \& {Carr}, B.~J. 1984, \apj, 280, 825

\bibitem[{{Bouret} {et~al.}(2012){Bouret}, {Hillier}, {Lanz}, \&
  {Fullerton}}]{Bouret12}
{Bouret}, J.-C., {Hillier}, D.~J., {Lanz}, T., \& {Fullerton}, A.~W. 2012,
  \aap, 544, A67

\bibitem[{{Brott} {et~al.}(2011){Brott}, {de Mink}, {Cantiello}, {Langer}, {de
  Koter}, {Evans}, {Hunter}, {Trundle}, \& {Vink}}]{Brott11}
{Brott}, I., {de Mink}, S.~E., {Cantiello}, M., {et~al.} 2011, \aap, 530, A115

\bibitem[{{Castor} {et~al.}(1975){Castor}, {Abbott}, \& {Klein}}]{Castor75}
{Castor}, J.~I., {Abbott}, D.~C., \& {Klein}, R.~I. 1975, \apj, 195, 157

\bibitem[{{Choi} {et~al.}(2016){Choi}, {Dotter}, {Conroy}, {Cantiello},
  {Paxton}, \& {Johnson}}]{Choi16}
{Choi}, J., {Dotter}, A., {Conroy}, C., {et~al.} 2016, \apj, 823, 102

\bibitem[{{Cohen} {et~al.}(2014){Cohen}, {Wollman}, {Leutenegger}, {Sundqvist},
  {Fullerton}, {Zsarg{\'o}}, \& {Owocki}}]{Cohen14}
{Cohen}, D.~H., {Wollman}, E.~E., {Leutenegger}, M.~A., {et~al.} 2014, \mnras,
  439, 908

\bibitem[{{Crowther} {et~al.}(2006){Crowther}, {Lennon}, \&
  {Walborn}}]{Crowther06}
{Crowther}, P.~A., {Lennon}, D.~J., \& {Walborn}, N.~R. 2006, \aap, 446, 279

\bibitem[{{de Jager} {et~al.}(1988){de Jager}, {Nieuwenhuijzen}, \& {van der
  Hucht}}]{deJager88}
{de Jager}, C., {Nieuwenhuijzen}, H., \& {van der Hucht}, K.~A. 1988, \aaps,
  72, 259

\bibitem[{{Dotter}(2016)}]{Dotter16}
{Dotter}, A. 2016, \apjs, 222, 8

\bibitem[{{Driessen} {et~al.}(2019){Driessen}, {Sundqvist}, \&
  {Kee}}]{Driessen19}
{Driessen}, F.~A., {Sundqvist}, J.~O., \& {Kee}, N.~D. 2019, \aap, 631, A172

\bibitem[{{Eggenberger} {et~al.}(2008){Eggenberger}, {Meynet}, {Maeder},
  {Hirschi}, {Charbonnel}, {Talon}, \& {Ekstr{\"o}m}}]{Eggenberger08}
{Eggenberger}, P., {Meynet}, G., {Maeder}, A., {et~al.} 2008, \apss, 316, 43

\bibitem[{{Fierlinger} {et~al.}(2016){Fierlinger}, {Burkert}, {Ntormousi},
  {Fierlinger}, {Schartmann}, {Ballone}, {Krause}, \& {Diehl}}]{Fierlinger16}
{Fierlinger}, K.~M., {Burkert}, A., {Ntormousi}, E., {et~al.} 2016, \mnras,
  456, 710

\bibitem[{{Fryer} \& {Kalogera}(2001)}]{Fryer01}
{Fryer}, C.~L. \& {Kalogera}, V. 2001, \apj, 554, 548

\bibitem[{{Glatzel} {et~al.}(1985){Glatzel}, {Fricke}, \& {El Eid}}]{Glatzel85}
{Glatzel}, W., {Fricke}, K.~J., \& {El Eid}, M.~F. 1985, \aap, 149, 413

\bibitem[{{Gormaz-Matamala} {et~al.}(2021){Gormaz-Matamala}, {Cur{\'e}},
  {Hillier}, {Najarro}, {Kub{\'a}tov{\'a}}, \& {Kub{\'a}t}}]{Gormaz21}
{Gormaz-Matamala}, A.~C., {Cur{\'e}}, M., {Hillier}, D.~J., {et~al.} 2021,
  \apj, 920, 64

\bibitem[{{Hawcroft} {et~al.}(2021){Hawcroft}, {Sana}, {Mahy}, {Sundqvist},
  {Abdul-Masih}, {Bouret}, {Brands}, {de Koter}, {Driessen}, \&
  {Puls}}]{Hawcroft21}
{Hawcroft}, C., {Sana}, H., {Mahy}, L., {et~al.} 2021, \aa, submitted

\bibitem[{{Heger} {et~al.}(2000){Heger}, {Langer}, \& {Woosley}}]{Heger00}
{Heger}, A., {Langer}, N., \& {Woosley}, S.~E. 2000, \apj, 528, 368

\bibitem[{{Heger} \& {Woosley}(2002)}]{Heger02}
{Heger}, A. \& {Woosley}, S.~E. 2002, \apj, 567, 532

\bibitem[{{Hillier} \& {Miller}(1998)}]{Hillier98}
{Hillier}, D.~J. \& {Miller}, D.~L. 1998, \apj, 496, 407

\bibitem[{{Howarth} {et~al.}(1997){Howarth}, {Siebert}, {Hussain}, \&
  {Prinja}}]{Howarth97}
{Howarth}, I.~D., {Siebert}, K.~W., {Hussain}, G.~A.~J., \& {Prinja}, R.~K.
  1997, \mnras, 284, 265

\bibitem[{{Jiang} {et~al.}(2018){Jiang}, {Cantiello}, {Bildsten}, {Quataert},
  {Blaes}, \& {Stone}}]{Jiang18}
{Jiang}, Y.-F., {Cantiello}, M., {Bildsten}, L., {et~al.} 2018, arXiv e-prints,
  arXiv:1809.10187

\bibitem[{{Kee} {et~al.}(2021){Kee}, {Sundqvist}, {Decin}, {de Koter}, \&
  {Sana}}]{Kee21}
{Kee}, N.~D., {Sundqvist}, J.~O., {Decin}, L., {de Koter}, A., \& {Sana}, H.
  2021, \aap, 646, A180

\bibitem[{{Keszthelyi} {et~al.}(2017){Keszthelyi}, {Puls}, \&
  {Wade}}]{Keszthelyi17}
{Keszthelyi}, Z., {Puls}, J., \& {Wade}, G.~A. 2017, \aap, 598, A4

\bibitem[{{Krti{\v c}ka} \& {Kub{\'a}t}(2010)}]{Krticka10}
{Krti{\v c}ka}, J. \& {Kub{\'a}t}, J. 2010, \aap, 519, A50

\bibitem[{{Krti{\v{c}}ka} \& {Kub{\'a}t}(2018)}]{Krticka18}
{Krti{\v{c}}ka}, J. \& {Kub{\'a}t}, J. 2018, \aap, 612, A20

\bibitem[{{Krti{\v{c}}ka} {et~al.}(2021){Krti{\v{c}}ka}, {Kub{\'a}t}, \&
  {Krti{\v{c}}kov{\'a}}}]{Krticka21}
{Krti{\v{c}}ka}, J., {Kub{\'a}t}, J., \& {Krti{\v{c}}kov{\'a}}, I. 2021, \aap,
  647, A28

\bibitem[{{Lagae} {et~al.}(2021){Lagae}, {Driessen}, {Hennicker}, {Kee}, \&
  {Sundqvist}}]{Lagae21}
{Lagae}, C., {Driessen}, F.~A., {Hennicker}, L., {Kee}, N.~D., \& {Sundqvist},
  J.~O. 2021, \aap, 648, A94

\bibitem[{{Langer}(1998)}]{Langer98}
{Langer}, N. 1998, \aap, 329, 551

\bibitem[{{Lucy}(1971)}]{Lucy71}
{Lucy}, L.~B. 1971, \apj, 163, 95

\bibitem[{{Maeder} \& {Meynet}(2000)}]{Maeder00}
{Maeder}, A. \& {Meynet}, G. 2000, \araa, 38, 143

\bibitem[{{Markova} \& {Puls}(2008)}]{Markova08}
{Markova}, N. \& {Puls}, J. 2008, \aap, 478, 823

\bibitem[{{Meynet} {et~al.}(1994){Meynet}, {Maeder}, {Schaller}, {Schaerer}, \&
  {Charbonnel}}]{Meynet94}
{Meynet}, G., {Maeder}, A., {Schaller}, G., {Schaerer}, D., \& {Charbonnel}, C.
  1994, \aaps, 103, 97

\bibitem[{{Muijres} {et~al.}(2011){Muijres}, {de Koter}, {Vink}, {Krti{\v
  c}ka}, {Kub{\'a}t}, \& {Langer}}]{Muijres11}
{Muijres}, L.~E., {de Koter}, A., {Vink}, J.~S., {et~al.} 2011, \aap, 526, A32

\bibitem[{{M{\"u}ller} \& {Vink}(2008)}]{Muller08}
{M{\"u}ller}, P.~E. \& {Vink}, J.~S. 2008, \aap, 492, 493

\bibitem[{{Najarro} {et~al.}(2011){Najarro}, {Hanson}, \& {Puls}}]{Najarro11}
{Najarro}, F., {Hanson}, M.~M., \& {Puls}, J. 2011, \aap, 535, A32

\bibitem[{{Nugis} \& {Lamers}(2000)}]{Nugis00}
{Nugis}, T. \& {Lamers}, H.~J.~G.~L.~M. 2000, \aap, 360, 227

\bibitem[{{Ober} {et~al.}(1983){Ober}, {El Eid}, \& {Fricke}}]{Ober83}
{Ober}, W.~W., {El Eid}, M.~F., \& {Fricke}, K.~J. 1983, \aap, 119, 61

\bibitem[{{Owocki}(2015)}]{Owocki15}
{Owocki}, S.~P. 2015, {Instabilities in the Envelopes and Winds of Very Massive
  Stars}, Vol. 412, 113

\bibitem[{{Paxton} {et~al.}(2011){Paxton}, {Bildsten}, {Dotter}, {Herwig},
  {Lesaffre}, \& {Timmes}}]{Paxton11}
{Paxton}, B., {Bildsten}, L., {Dotter}, A., {et~al.} 2011, \apjs, 192, 3

\bibitem[{{Paxton} {et~al.}(2013){Paxton}, {Cantiello}, {Arras}, {Bildsten},
  {Brown}, {Dotter}, {Mankovich}, {Montgomery}, {Stello}, {Timmes}, \&
  {Townsend}}]{Paxton13}
{Paxton}, B., {Cantiello}, M., {Arras}, P., {et~al.} 2013, \apjs, 208, 4

\bibitem[{{Paxton} {et~al.}(2015){Paxton}, {Marchant}, {Schwab}, {Bauer},
  {Bildsten}, {Cantiello}, {Dessart}, {Farmer}, {Hu}, {Langer}, {Townsend},
  {Townsley}, \& {Timmes}}]{Paxton15}
{Paxton}, B., {Marchant}, P., {Schwab}, J., {et~al.} 2015, \apjs, 220, 15

\bibitem[{{Paxton} {et~al.}(2018){Paxton}, {Schwab}, {Bauer}, {Bildsten},
  {Blinnikov}, {Duffell}, {Farmer}, {Goldberg}, {Marchant}, {Sorokina},
  {Thoul}, {Townsend}, \& {Timmes}}]{Paxton18}
{Paxton}, B., {Schwab}, J., {Bauer}, E.~B., {et~al.} 2018, \apjs, 234, 34

\bibitem[{{Paxton} {et~al.}(2019){Paxton}, {Smolec}, {Schwab}, {Gautschy},
  {Bildsten}, {Cantiello}, {Dotter}, {Farmer}, {Goldberg}, {Jermyn}, {Kanbur},
  {Marchant}, {Thoul}, {Townsend}, {Wolf}, {Zhang}, \& {Timmes}}]{Paxton19}
{Paxton}, B., {Smolec}, R., {Schwab}, J., {et~al.} 2019, \apjs, 243, 10

\bibitem[{{Petrov} {et~al.}(2016){Petrov}, {Vink}, \&
  {Gr{\"a}fener}}]{Petrov16}
{Petrov}, B., {Vink}, J.~S., \& {Gr{\"a}fener}, G. 2016, \mnras, 458, 1999

\bibitem[{{Poniatowski} {et~al.}(2021){Poniatowski}, {Sundqvist}, {Kee},
  {Owocki}, {Marchant}, {Decin}, {de Koter}, {Mahy}, \& {Sana}}]{Poniatowski21}
{Poniatowski}, L.~G., {Sundqvist}, J.~O., {Kee}, N.~D., {et~al.} 2021, \aap,
  647, A151

\bibitem[{{Puls} {et~al.}(2020){Puls}, {Najarro}, {Sundqvist}, \&
  {Sen}}]{Puls20}
{Puls}, J., {Najarro}, F., {Sundqvist}, J.~O., \& {Sen}, K. 2020, \aap, 642,
  A172

\bibitem[{{Puls} {et~al.}(2008){Puls}, {Vink}, \& {Najarro}}]{Puls08}
{Puls}, J., {Vink}, J.~S., \& {Najarro}, F. 2008, \aapr, 16, 209

\bibitem[{{Reimers}(1975)}]{Reimers75}
{Reimers}, D. 1975, Memoires of the Societe Royale des Sciences de Liege, 8,
  369

\bibitem[{{Rubio-D\'iez} {et~al.}(2021){Rubio-D\'iez}, {Sundqvist}, {Najarro},
  {Traficante}, {Puls}, {Calzoletti}, \& {Figer}}]{RubioDiez21}
{Rubio-D\'iez}, M.~M., {Sundqvist}, J.~O., {Najarro}, F., {et~al.} 2021,
  accepted by A\&A

\bibitem[{{Sander} {et~al.}(2017){Sander}, {Hamann}, {Todt}, {Hainich}, \&
  {Shenar}}]{Sander17}
{Sander}, A.~A.~C., {Hamann}, W.-R., {Todt}, H., {Hainich}, R., \& {Shenar}, T.
  2017, \aap, 603, A86

\bibitem[{{Sander} {et~al.}(2020){Sander}, {Vink}, \& {Hamann}}]{Sander20}
{Sander}, A. A.~C., {Vink}, J.~S., \& {Hamann}, W.~R. 2020, \mnras, 491, 4406

\bibitem[{{Schultz} {et~al.}(2020){Schultz}, {Bildsten}, \&
  {Jiang}}]{Schultz20}
{Schultz}, W.~C., {Bildsten}, L., \& {Jiang}, Y.-F. 2020, \apj, 902, 67

\bibitem[{{Shenar} {et~al.}(2020){Shenar}, {Gilkis}, {Vink}, {Sana}, \&
  {Sander}}]{Shenar20}
{Shenar}, T., {Gilkis}, A., {Vink}, J.~S., {Sana}, H., \& {Sander}, A.~A.~C.
  2020, \aap, 634, A79

\bibitem[{{Shenar} {et~al.}(2016){Shenar}, {Hainich}, {Todt}, {Sander},
  {Hamann}, {Moffat}, {Eldridge}, {Pablo}, {Oskinova}, \&
  {Richardson}}]{Shenar16}
{Shenar}, T., {Hainich}, R., {Todt}, H., {et~al.} 2016, \aap, 591, A22

\bibitem[{{Shenar} {et~al.}(2015){Shenar}, {Oskinova}, {Hamann}, {Corcoran},
  {Moffat}, {Pablo}, {Richardson}, {Waldron}, {Huenemoerder}, {Ma{\'\i}z
  Apell{\'a}niz}, {Nichols}, {Todt}, {Naz{\'e}}, {Hoffman}, {Pollock}, \&
  {Negueruela}}]{Shenar15}
{Shenar}, T., {Oskinova}, L., {Hamann}, W.~R., {et~al.} 2015, \apj, 809, 135

\bibitem[{{Sim{\'o}n-D{\'\i}az} \& {Herrero}(2014)}]{Simon-Diaz14}
{Sim{\'o}n-D{\'\i}az}, S. \& {Herrero}, A. 2014, \aap, 562, A135

\bibitem[{{Smith}(2014)}]{Smith14}
{Smith}, N. 2014, \araa, 52, 487

\bibitem[{{Sundqvist} {et~al.}(2019){Sundqvist}, {Bj{\"o}rklund}, {Puls}, \&
  {Najarro}}]{Sundqvist19}
{Sundqvist}, J.~O., {Bj{\"o}rklund}, R., {Puls}, J., \& {Najarro}, F. 2019,
  \aap, 632, A126

\bibitem[{{Sundqvist} {et~al.}(2011){Sundqvist}, {Puls}, {Feldmeier}, \&
  {Owocki}}]{Sundqvist11}
{Sundqvist}, J.~O., {Puls}, J., {Feldmeier}, A., \& {Owocki}, S.~P. 2011, \aap,
  528, 64

\bibitem[{{Sundqvist} {et~al.}(2014){Sundqvist}, {Puls}, \&
  {Owocki}}]{Sundqvist14}
{Sundqvist}, J.~O., {Puls}, J., \& {Owocki}, S.~P. 2014, \aap, 568, 59

\bibitem[{{{\v S}urlan} {et~al.}(2013){{\v S}urlan}, {Hamann}, {Aret},
  {Kub{\'a}t}, {Oskinova}, \& {Torres}}]{Surlan13}
{{\v S}urlan}, B., {Hamann}, W.-R., {Aret}, A., {et~al.} 2013, \aap, 559, A130

\bibitem[{{Vink} {et~al.}(1999){Vink}, {de Koter}, \& {Lamers}}]{Vink99}
{Vink}, J.~S., {de Koter}, A., \& {Lamers}, H.~J.~G.~L.~M. 1999, \aap, 350, 181

\bibitem[{{Vink} {et~al.}(2000){Vink}, {de Koter}, \& {Lamers}}]{Vink00}
{Vink}, J.~S., {de Koter}, A., \& {Lamers}, H.~J.~G.~L.~M. 2000, \aap, 362, 295

\bibitem[{{Vink} {et~al.}(2001){Vink}, {de Koter}, \& {Lamers}}]{Vink01}
{Vink}, J.~S., {de Koter}, A., \& {Lamers}, H.~J.~G.~L.~M. 2001, \aap, 369, 574

\bibitem[{{Vink} {et~al.}(2021){Vink}, {Higgins}, {Sander}, \&
  {Sabhahit}}]{Vink21a}
{Vink}, J.~S., {Higgins}, E.~R., {Sander}, A. A.~C., \& {Sabhahit}, G.~N. 2021,
  \mnras, 504, 146

\bibitem[{{Vink} \& {Sander}(2021)}]{Vink21b}
{Vink}, J.~S. \& {Sander}, A. A.~C. 2021, \mnras, 504, 2051

\bibitem[{{Woosley}(2017)}]{Woosley17}
{Woosley}, S.~E. 2017, \apj, 836, 244

\bibitem[{{Woosley} {et~al.}(2007){Woosley}, {Blinnikov}, \&
  {Heger}}]{Woosley07}
{Woosley}, S.~E., {Blinnikov}, S., \& {Heger}, A. 2007, \nat, 450, 390

\end{thebibliography}

\appendix
\section{Additional figures}\label{appendix}
\begin{figure*}
        \centering
        \includegraphics[width=0.49\hsize]{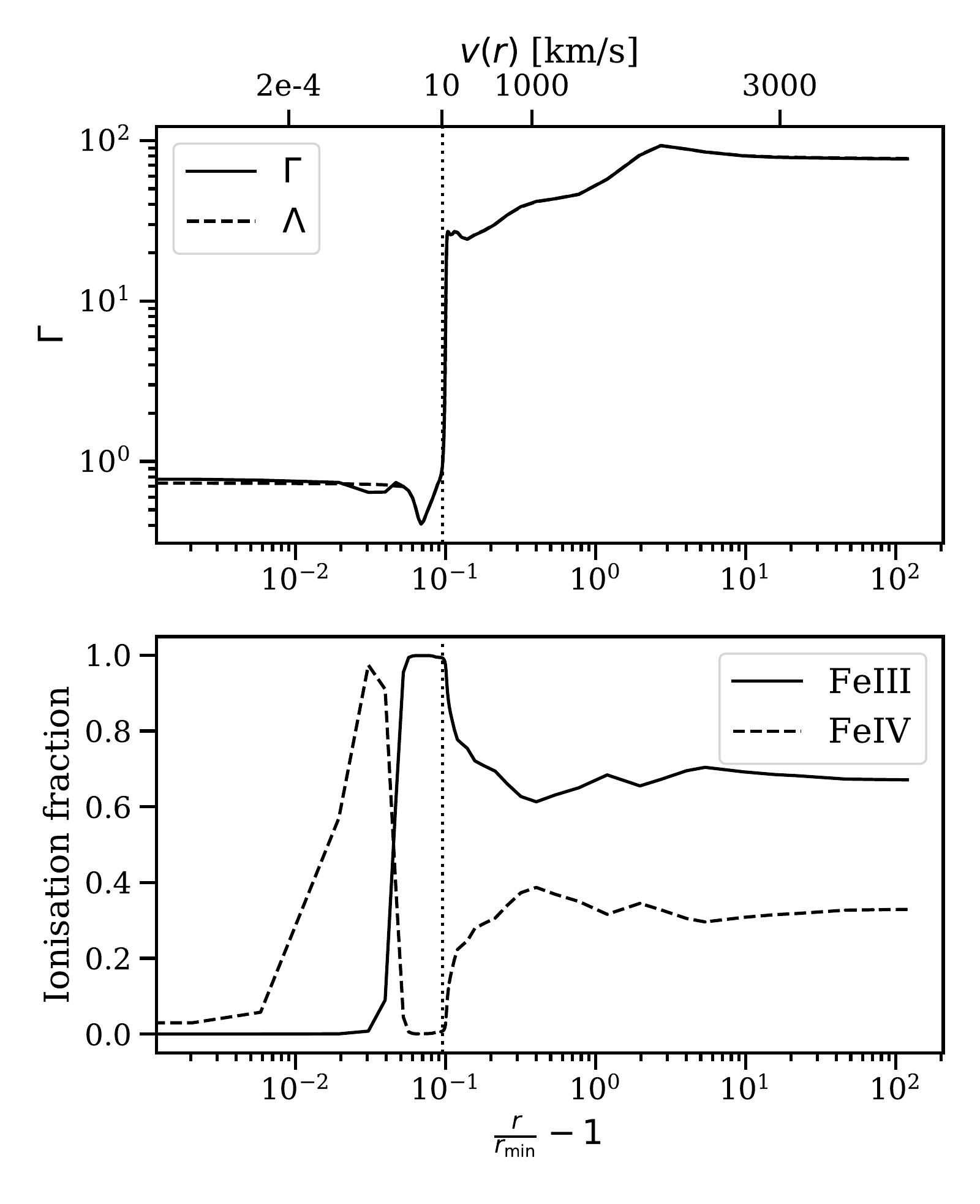}
        \includegraphics[width=0.49\hsize]{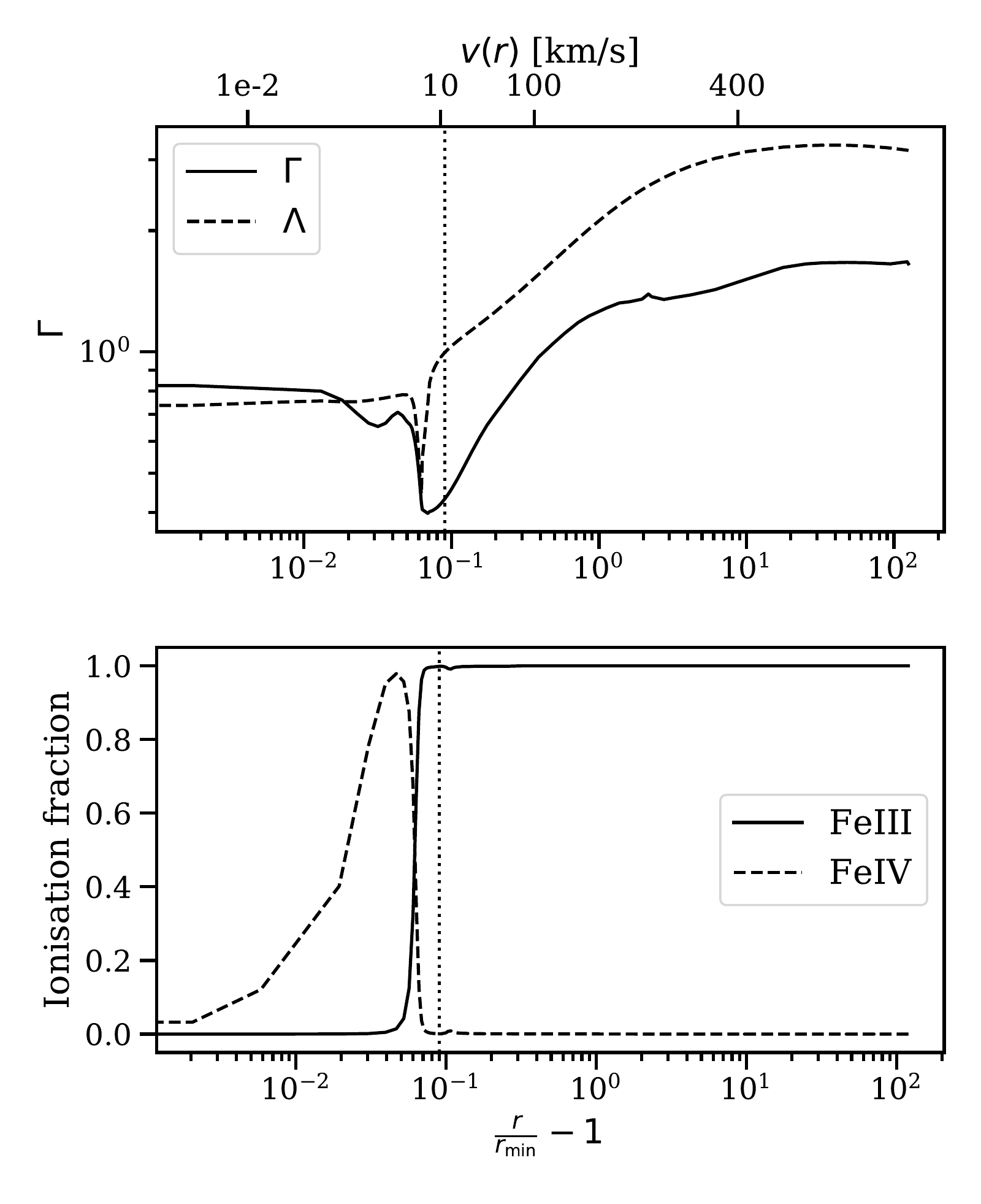}
        \caption{Same figure as Fig. \ref{L45_bistab} but now with $\log(L_{\ast}/\Lsol) = 5.75$, $M_{\ast} = 50 \Msol$, and $\Teff=15000$K. The self-consistent model has $\Mdot=7.67\cdot10^{-8}$ \Msol/yr and $\vinf=3111$ km/s, while the $\beta$-model has $\Mdot=5.38\cdot10^{-6}$ \Msol/yr and $\vinf=451$ km/s.}
        \label{L575_bistab}
    \end{figure*}

    Fig. \ref{L575_bistab} shows a similar comparison as in Fig. \ref{L45_bistab} of a self-consistent model and a model using a $\beta$-velocity law of $\beta=1$, but now at higher mass and luminosity. The mass-loss rate of the $\beta$-model is about 70 times higher than the dynamically-consistent value. This model then again shows a mismatch between $\Gamma$ and $\Lambda$ with a value of $\Gamma=0.45$ at the sonic points, which is significantly below the required value $\approx1$.

\section{Comparison with $\Gamma$ from CMFGEN} \label{check_grad}

\begin{table*}
\caption{Stellar and wind parameters of our model grid used to check the radiative acceleration in the B super-/hypergiant domain.  All models have a luminosity of $\log L_{\ast}/\Lsol = 5.76$, and have been calculated with a velocity field exponent $\beta = 1$, a micro-turbulent velocity $\vturb = 10~\kms$, an unclumped wind, no X-ray emission from wind-embedded shocks, and solar abundances from \citet{asplund09}, in particular \YHe =$N_{\rm He}/N_{\rm H}$ = 0.1.}
\label{tabgrid}
\tabcolsep1.5mm
\begin{center}
\begin{tabular}{lccrccrc}
\hline 
\hline
\multicolumn{1}{c}{Model}
&\multicolumn{1}{c}{Mass}
&\multicolumn{1}{c}{\Teff}
&\multicolumn{1}{c}{\Rstar}
&\multicolumn{1}{c}{\logg}
&\multicolumn{1}{c}{$\Gamma_{\rm e}$}
&\multicolumn{1}{c}{\Mdot}
&\multicolumn{1}{c}{\vinf}\\
\multicolumn{1}{c}{}
&\multicolumn{1}{c}{\Msol}
&\multicolumn{1}{c}{(K)}
&\multicolumn{1}{c}{(\Rsol)}
&\multicolumn{1}{c}{(cgs)}
&\multicolumn{1}{c}{}
&\multicolumn{1}{c}{(\Msol {\rm yr}$^{-1}$)}
&\multicolumn{1}{c}{(\kms)}
\\
\hline
B225\_1 & 31.5 & 22500 & 49.3 & 2.55 & 0.43 &  8.20$\cdot 10^{-6}$  & 678\\
B225\_2 & 31.5 & 22500 & 49.3 & 2.55 & 0.43 &  0.73$\cdot 10^{-6}$  & 678\\
B200\_1 & 30.1 & 20000 & 62.4 & 2.33 & 0.45 & 11.20$\cdot 10^{-6}$  & 556\\
B200\_2 & 50.5 & 20000 & 62.4 & 2.55 & 0.27 &  3.70$\cdot 10^{-6}$  & 560\\
B175\_1 & 33.0 & 17500 & 71.3 & 2.25 & 0.31 & 14.30$\cdot 10^{-6}$  & 397\\
B175\_2 & 33.0 & 17500 & 71.3 & 2.25 & 0.31 &  3.00$\cdot 10^{-6}$  & 430\\
B150\_1 & 37.8 & 15000 & 83.2 & 2.17 & 0.20 & 14.30$\cdot 10^{-6}$  & 397\\
B150\_2 & 45.0 & 15000 & 83.2 & 2.25 & 0.17 &  3.50$\cdot 10^{-6}$  & 430\\
\hline
\end{tabular}
\end{center}
\end{table*}

\begin{figure*}
\resizebox{\hsize}{!}
  {\includegraphics[angle=90]{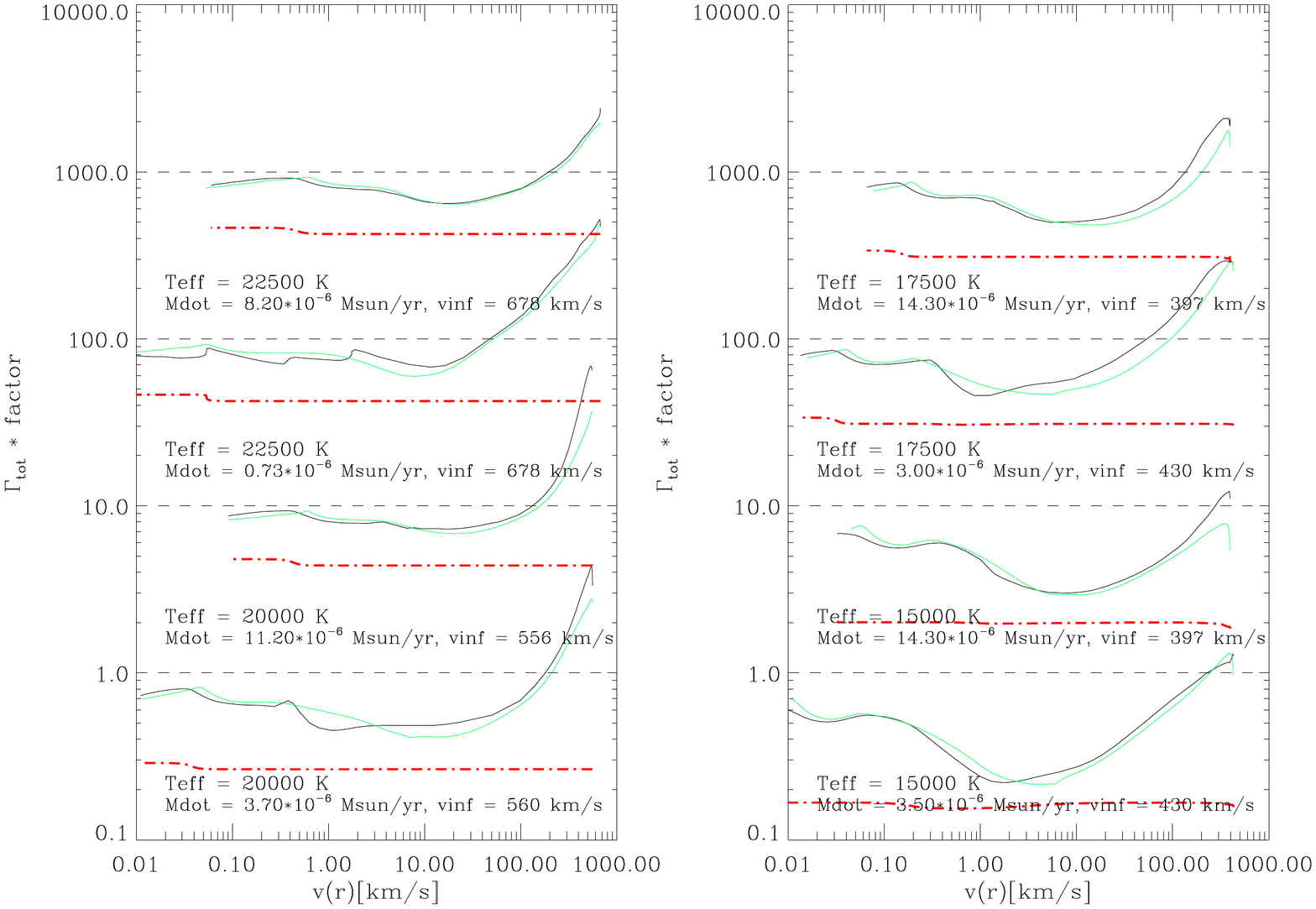}} 
\caption{$\Gamma_{\rm tot} = g_{\rm rad}^{\rm tot}/g_{\rm grav}$ for all models from our grid (see Table~\ref{tabgrid}), as a function of velocity.  Black: results from {\sc fastwind} v11, green: results from {\sc cmfgen}. To improve the visibility, all $\Gamma$-values have been multiplied with factors $10^i, i \in [0,4]$, from bottom to top. The red dashed-dotted lines correspond to $\Gamma_{\rm e}(r)$ for pure electron scattering (its local variation relates to changes in the ionization structure, mostly of H and He), and the dashed lines indicate, for each model, the relation $\Gamma_{\rm tot} = 1$. We note that all models displayed here have been calculated with a fixed\, $\beta=1$ velocity law, i.e., they are {\textit not} self-consistent. See text.} 
\label{comp_grad}
\end{figure*}
The self-consistent models presented and discussed in the main part of this paper show significant differences in their wind-structure (both with respect to mass-loss rate and velocity law) compared to previous predictions (e.g., from \citealt{Vink00, Vink01}), particularly in the B-star range. To underpin these results, we need to convince ourselves that the radiative acceleration as calculated by {\sc fastwind} does not suffer from specific problems, especially from deficiencies in the used line-lists resulting in a predicted line-acceleration that is too low. To this end, we proceed as in \citet{Puls20}, who compared the radiative acceleration in the O/early B-star regime (down to \Teff $\approx$ 25~kK) with corresponding results from {\sc cmfgen} models. The latter code has been used, e.g., also in theoretical investigations about the location of the potential bi-stability region preformed by \citet{Petrov16}. 

We set up a small model grid enclosing the critical region around \Teff\ $\approx$ 20~kK and concentrating on B super-/hypergiants, assuming a prototypical $\beta = 1$ velocity law as also used in the calculations by \citet{Vink00, Vink01}. We neglected wind-clumping and X-ray emission from wind-embedded shocks. The division between the $\beta$ law and the quasi-hydrostatic stratification was set (in most cases) to 0.5 $\varv_{\rm sound}$. For each model we considered two mass-loss rates, a high one following the trend predicted with the mass-loss recipes by \citet{Vink00,Vink01} and a lower one to check the influence of a less dense wind. All parameters of the grid-stars are provided in Table~\ref{tabgrid}.

The outcome of our modelling is displayed in Fig.~\ref{comp_grad}, where we compare the total radiative acceleration as calculated by {\sc fastwind} v11 and {\sc cmfgen} as a function of velocity, measured in units of gravitational acceleration (i.e., we display $\Gamma$ as defined in Eq. 1). The red dashed-dotted lines indicate the conventional Gamma-factor for electron scattering, $\Gamma_{\rm e}$, as defined in the beginning of Sect. 4, also as a function of velocity. 

The comparison shows clearly that in most cases the photospheric accelerations (even in those cases where $\Gamma$ is close to unity and thus has a significant effect on the structure) agree quite well, and that also the predicted wind acceleration (at least until 0.5\vinf) does not show significant differences (in all cases, below 20\%). Only for the models at 20 and 17.5 kK, {\sc fastwind} predicts a larger acceleration in the outer wind, by a factor 1.5 to 2. Though the origin of this discrepancy is currently unclear (work in progress), our modelling shows that, at least compared to {\sc cmfgen}, there is no deficiency in the line-force which might have influenced our results from the previous sections. Particularly in the most important region around the sonic point (where \Mdot\ is determined) until few hundreds of \kms\ (a region which also controls the lowermost wind-regime, due to backscattering), the agreement between the two independent codes and predictions is reasonable (often even better).

Given that the two codes use a rather different computational approach and different atomic data bases, the outcome of our comparison is ensuring, and we are confident that our calculations presented in the main part of this paper are not affected by too low $\grad$ because of missing lines. As already noted in previous publications of this series, and also by \citet{Puls20} for the O-star case, also here $\Gamma_{\rm tot} = 1$ is reached only at substantial velocities (100~\kms or more), whereas, in self-consistent wind-models, this should happen very close to the sonic point (here: 15 {\ldots} 18 \kms, in dependence of \Teff). This highlights that the displayed models are far away from being hydrodynamically self-consistent. To achieve such a self-consistency, and due to the basic dependence $\grad \propto (\dd \varv/\dd r)/\rho$ (e.g., \citealt{Castor75}), only a steeper velocity law and a lower mass-loss rate will lead to a higher \grad\ in the transonic region, which is also the basic outcome of our self-consistent calculations.
\end{document}